\documentclass[12pt]{article}
\usepackage{epsfig}

\catcode`@=12
\newcommand{\be} {\begin{equation}}
\newcommand{\ee} {\end{equation}}
\newcommand{\bea}{\begin{eqnarray}}
\newcommand{\eea}{\end{eqnarray}}
\newcommand{\ff}  {\begin{flushleft}}
\newcommand{\ef} {\end{flushleft}}
\def\n{\noindent}

\def\s12{\sin\theta_{12}}
\def\s23{\sin\theta_{23}}
\def\s13{\sin\theta_{13}}
\def\ts12{\theta_{12}}
\def\e{\epsilon}
\makeatletter
\@addtoreset{equation}{section}

\makeatother
 \title{\LARGE\bf Baryon asymmetry from leptogenesis with four zero
neutrino Yukawa textures}
\author{{\bf Biswajit Adhikary$^{\rm a, b}$\footnote{biswajit.adhikary
@saha.ac.in}
     , Ambar Ghosal$^{\rm a}$\footnote{ambar.ghosal@saha.ac.in} and
Probir Roy$^{\rm a}$\footnote{probir.roy@saha.ac.in}}\\
  a) Saha Institute of Nuclear Physics, 1/AF Bidhannagar,
  Kolkata 700064, India \\
  b)Department of Physics, Gurudas College,
Narkeldanga,
 Kolkata-700054, India}
\begin{document}
\maketitle
\thispagestyle{empty}
\begin{abstract}
{
The generation of the right amount of baryon asymmetry $\eta$ 
of the Universe from supersymmetric leptogenesis is studied within 
the type-I seesaw framework with three heavy singlet Majorana neutrinos 
$N_i\,\,(i = 1,2,3)$ and their superpartners. We assume  the occurrence 
of four zeroes in the neutrino 
Yukawa coupling matrix $Y_\nu$, taken to be 
$\mu\tau$ symmetric, in the weak 
basis where $N_i$ (with real masses $M_i>0$) and the charged leptons 
$l_\alpha\,\, (\alpha = e,\mu,\tau)$ are mass diagonal. The quadrant of the 
single nontrivial phase, allowed in the corresponding light 
neutrino mass matrix 
$m_\nu$, gets fixed and additional constraints ensue from the requirement 
of matching $\eta$ with its observed value. Special attention is paid to 
flavor effects in the washout of the lepton asymmetry. We also comment on 
the role of small departures
from high scale $\mu\tau$ symmetry due to RG evolution. 
}
\end{abstract}
\section{Introduction}
Baryogenesis through leptogenesis \cite{one, one2, khlopov} 
is a simple and attractive 
mechanism to explain the mysterious excess of matter over antimatter in 
the Universe. A lepton asymmetry is first generated at a relatively high scale 
$(>10^9\,\,\rm{GeV})$. This then gets converted into a nonzero $\eta$, the 
difference between the baryonic and antibaryonic number densities normalized 
to the photon number density $(n_B - n_{\bar B}){n^{-1}_\gamma}$, 
at electroweak 
temperatures \cite{two} due to $B+L$ violating but $B-L$ conserving sphaleron 
interactions of the Standard Model. Since the origin of the lepton asymmetry 
is from 
 out of equilibrium decays of 
heavy unstable singlet Majorana neutrinos \cite{three}, 
the type-I seesaw framework 
\cite{four, four2, four3, four4, valle}, proposed for the generation of 
light neutrino masses, is ideal for this purpose. We study baryogenesis 
via supersymmetric leptogenesis \cite{giu} 
with a type-I seesaw driven by three heavy $(>10^9\,\,\rm{GeV})$ 
right-chiral Majorana neutrinos $N_i\,\,(i=1,2,3)$ with Yukawa 
couplings to the known left chiral neutrinos through the 
relevant Higgs doublet. 
There have been some recent investigations 
\cite{add, asaka, canetti, Adhikary:2008au} 
studying the 
interrelation between leptogenesis, heavy right-chiral neutrinos 
and neutrino flavor mixing. However, our angle is a little bit 
different in that we link supersymmetric leptogenesis to zeroes in the neutrino Yukawa coupling matrix. In fact, we take a $\mu\tau$ symmetric \cite{five} 
neutrino Yukawa coupling matrix $Y_\nu$ with four 
zeroes \cite{six} in the weak basis specified in the abstract.
\vskip 0.1in
\noindent
There are several reasons for our choice. First, a seesaw with three 
heavy right chiral neutrinos is the simplest type-I scheme yielding 
a square Yukawa coupling matrix $Y_\nu$ on which symmetries can be 
imposed in a straightforward way. Second, $\mu\tau$ symmetry 
\cite{mutau} - \cite{Ghosal}
in the neutrino sector provides a very natural way of understanding the 
observed maximal mixing of atmospheric neutrinos. Though it also 
predicts a vanishing value for the neutrino mixing angle $\theta_{13}$, 
the latter is known from reactor experiments to be rather small. A tiny 
nonzero value of $\theta_{13}$ could arise at the 1-loop level via 
the charged 
lepton sector, where $\mu\tau$ symmetry is obviously broken, though RG 
effects if the said symmetry is imposed at a high scale \cite{five}. Third, 
four has been shown \cite{six} to be the maximum number of zeroes 
phenomenologically allowed in $Y_\nu$ within the type-I seesaw framework in 
the weak basis described earlier. Finally, four zero neutrino Yukawa 
textures provide \cite{eight} a very constrained and predictive theoretical 
scheme - particularly if $\mu\tau$ symmetry is imposed \cite{five}.
\vskip 0.1in
\noindent
The beautiful thing about such four zero textures in $Y_\nu$ is that the high 
scale CP violation, required for leptogenesis, gets completely specified here
\cite{six} in terms of CP violation that is observable in the laboratory 
with neutrino and antineutrino beams. In our $\mu\tau$ symmetric scheme 
\cite{five}, which admits two categories $A$ and $B$, 
the latter is given in terms 
of just one phase (for each category) which is already quite 
constrained by the extant neutrino oscillation data. 
Indeed, the quadrant in which this phase lies - 
which was earlier unspecified by the same data - gets 
fixed by the requirement of generating the right size and sign of 
the baryon asymmetry. Moreover, the magnitude 
of this phase is further constrained.
\vskip 0.1in
\noindent                 
In computing the net lepton asymmetry generated at a high scale,  
one needs to consider not only the 
decays of heavy right-chiral neutrinos $N_i$ into Higgs and left-chiral 
lepton doublets as well as their superpartner versions 
but also the washout caused by inverse decay processes in 
the thermal bath. The 
role of flavor \cite{bar, abada1, nardi, thirteen} 
can be crucial in the latter. In the Minimal Supersymmetric 
Standard Model (MSSM \cite{nine}), this has been studied \cite{ten} through 
flavor dependent Boltzmann equations. The solutions to those equations 
demonstrate that flavor effects show up differently in three distinct regimes 
depending on the mass of the lightest of the three heavy neutrinos and an 
MSSM parameter $\tan\beta$ which is the ratio $v_u/v_d$ of the up-type 
and down-type Higgs VEVs. In each regime there are three $N_i$ mass 
hierarchical cases : (a) normal, (b) inverted and (c) quasidegenerate. All 
these, considered in both categories $A$ and $B$, make up eighteen different 
possibilities 
for each of which the lepton asymmetry is calculated here. That then is 
converted into the baryon asymmetry by standard sphaleronic 
conversion and 
compared with observation. These lead to the phase constraints mentioned 
above as well as a stronger restriction on the parameter $\tan\beta$ in some 
cases.
\vskip 0.1in
\n
If $\mu\tau$ symmetry is posited at a high scale characterized by the masses of 
the heavy Majorana neutrinos, renormalization group evolution 
down to a laboratory energy $\lambda$ breaks it radiatively. 
Consequently, a small nonzero $\theta_{13}^\lambda$, crucially dependent 
on the magnitude of $\tan\beta$, gets induced. The said new restrictions 
on $\tan\beta$ coming from $\eta$ in some cases therefore cause strong
constraints on the nonzero value of $\theta_{13}^\lambda$ which we enumerate.
\vskip 0.1in
\noindent
One possible problem with high scale supersymmetric thermal 
leptogenesis is that 
of the overabundance of gravitinos caused by 
the high reheating temperature. For a decaying gravitino, this can lead 
to a  conflict with Big Bang Nucleosynthesis constraints,
while for a stable gravitino (dark matter) this poses 
the danger of overclosing the Universe.
The problem can be evaded by appropriate mass and lifetime restrictions on the 
concerned sparticles, cf. sec. 16.4 of ref \cite{nine}. 
Such is the case, for instance, with gauge mediated supersymmetry
breaking with a gravitino as light as $O$(KeV) in mass. In 
gravity mediated supersymmetry breaking there are sparticle mass regions
where the problem can be avoided -- especially within an 
inflationary scenario. An illustration is a model \cite{covi}, with
a gluino and a neutralino that are close in mass, which 
satisfies the BBN constraints. Purely cosmological solutions within 
the supersymmetric inflationary scenario have also been proposed, e.g. 
\cite{sanchez}.
We feel that, while the gravitino issue is one of concern,
 it can be resolved and therefore need not be addressed here any further.
 \vskip 0.1in
\noindent  
The plan of the rest of the paper is as follows. In section 2 we recount the 
properties of the allowed $\mu\tau$ symmetric four zero $Y_\nu$ 
textures. Section 3 contains an outline of the basic steps in our calculation
of $\eta$. In section 4, $\eta$ is computed in our scheme 
for the three different heavy neutrino 
mass hierarchical cases in the regimes of unflavoured, fully flavored 
and $\tau$-flavored leptogenesis for both categories $A$ and $B$. 
Section 5 consists of our results on constraints emerging 
from $\eta$ on the allowed $\mu\tau$ symmetric four zero $Y_\nu$ 
textures. In section 6 we discuss 
the departures - due to RG evolution down to laboratory 
energies - from $\mu\tau$ symmetry imposed at a high scale 
$\sim$ ${\rm min}\,\,(M_1, M_2, M_3)\equiv M_{lowest}$. 
Section 7 summarizes our conclusions. Appendices A, B and C
list the detailed expressions for $\eta$ in each of the eighteen different 
possibilities.
\section{Allowed $\mu\tau$ symmetric four zero textures of $Y_\nu$}
\vskip 0.1in
\noindent
The complex symmetric light neutrino Majorana mass matrix $m_\nu$ is given 
in our basis by
\be
m_\nu = -\frac{1}{2} v_u^2 Y_\nu {\rm diag}.(
M_1^{-1}, M_2^{-1}, M_3^{-1})Y_\nu^T
= U {\rm diag}.(m_1,m_2,m_3)U^T.
\ee    
We work within the confines of the MSSM \cite{nine} so that 
$v_u = v\sin\beta$ and the W-mass equals $\frac{1}{2}gv$, $g$ being the 
$SU(2)_L$ semiweak gauge coupling strength. The unitary PMNS mixing matrix  
 $U$ is parametrized as 
\be
U = \pmatrix{1&0&0\cr
             0&c_{23}&-s_{23}\cr
             0&s_{23}&c_{23}} 
     \pmatrix{c_{13}&0&-s_{13}e^{-i\delta_D}\cr
              0 & 1& 0\cr
              s_{13}e^{i\delta_D}&0&c_{13}}
     \pmatrix{c_{12}&s_{12}&0\cr
              -s_{12}&c_{12}&0\cr
              0 &0 &1}
     \pmatrix{e^{i\alpha_M}&0&0\cr
               0&e^{i\beta_M}&0\cr
               0&0&1},
\ee
\noindent
where $c_{ij}= \cos\theta_{ij}$, $s_{ij}=\sin\theta_{ij}$ and $\delta_D$, 
$\alpha_M$, $\beta_M$ are the Dirac phase and two Majorana phases 
respectively.
\vskip 0.1in
\noindent
The statement of $\mu\tau$ symmetry is that all couplings and masses in the 
pure neutrino part of the Lagrangian are invariant under the interchange 
of the flavor indices 2 and 3. Thus
\vskip 0.1in
\noindent
$$
{(Y_\nu)}_{12} = {(Y_\nu)}_{13},   
\eqno(2.3a)
$$
$$
{(Y_\nu)}_{21} = {(Y_\nu)}_{31},
\eqno(2.3b)
$$
$$
{(Y_\nu)}_{23} = {(Y_\nu)}_{32},
\eqno(2.3c)
$$
$$
{(Y_\nu)}_{22} = {(Y_\nu)}_{33}
\eqno(2.3d)
$$
and 
$$
M_2 = M_3.
\eqno(2.4)
$$
\noindent
Eqs. (2.3) and (2.4), in conjunction with eq.(2.1), lead to a custodial 
$\mu\tau$ symmetry in $m_\nu$ : 
$$
{(m_\nu)}_{12} = {(m_\nu)}_{21} = {(m_\nu)}_{13} =  {(m_\nu)}_{31},
\eqno(2.5a)
$$
$$
{(m_\nu)}_{22} = {(m_\nu)}_{33}.
\eqno(2.5b)
$$
\noindent
Eqs. (2.5) immediately imply that $\theta_{23} = \pi/4$ and $\theta_{13}=0$. 
With this $\mu\tau$ symmetry, it was shown in Ref. \cite{five} that only 
four textures 
with four zeroes in $Y_\nu$ are allowed. 
These fall into two categories $A$ and $B$ - each category containing a 
pair of textures yielding an identical form 
of $m_\nu$. These allowed textures may be written in the form of the Dirac 
mass matrix $m_D = Y_\nu v_u/\sqrt{2}$ in terms of complex parameters 
$a_1$, $a_2$, $b_1$, $b_2$. 
$$
{\it Category}\,\, A : m_{DA}^{(1)} = \pmatrix{a_1&a_2&a_2\cr
                                     0&0&b_1\cr
                                     0&b_1&0}, 
             m_{DA}^{(2)} = \pmatrix{a_1&a_2&a_2\cr
                                     0&b_1&0\cr
                                     0&0&b_1},
\eqno(2.6a)
$$
$$
{\it Category}\,\, B : m_{DB}^{(1)} = \pmatrix{a_1&0&0\cr
                                     b_1&0&b_2\cr
                                     b_1&b_2&0}, 
             m_{DB}^{(2)} = \pmatrix{a_1&0&0\cr
                                     b_1&b_2&0\cr
                                     b_1&0&b_2},
\eqno(2.6b)
$$
\noindent
The corresponding expressions for $m_\nu$, obtained via eq.(2.1), are 
much simplified by a change of variables. We introduce overall mass scales 
$m_{A,B}$, real parameters $k_1$, $k_2$, $l_1$, $l_2$ and phases $\bar\alpha$
and $\bar\beta$ defined by
\vskip 0.1in 
\noindent
{{\it Category}\,\, $A$} : 
$$
m_A = -b_1^2/M_2,\quad\,\, 
k_1={\left|{\frac{a_1}{b_1}}\right|}
\sqrt{\frac{M_2}{M_1}},\,\,\quad k_2 = \left|{\frac{a_2}{b_1}}\right|, 
\quad\,\,\bar{\alpha} = {\rm arg}
\frac{a_1}{a_2}.\quad\quad\quad\quad\quad\quad
\eqno(2.7a)
$$
\noindent
{{\it Category}\,\, $B$} : 
$$
m_B = -b_2^2/M_2, \quad l_1=\left|{\frac{a_1}{b_2}}\right|
\sqrt{\frac{M_2}{M_1}},\quad l_2 = \left|{\frac{b_1}{b_2}}\right|
\sqrt{\frac{M_2}{M_1}},\quad
\bar{\beta} = {\rm arg}\frac{b_1}{b_2}.
\eqno(2.7b)
$$
\noindent
Then the light neutrino mass matrix for each category can be written as 
\cite{ten}
$$
m_{\nu A} = m_A\pmatrix{k_1^2e^{2i\bar\alpha}+2k_2^2&k_2&k_2\cr
                        k_2 &1& 0\cr
                        k_2&0&1},
m_{\nu B} = m_B \pmatrix{l_1^2&l_1l_2e^{i\bar\beta}&l_1l_2e^{i\bar\beta}\cr
                                    l_1l_2e^{i\bar\beta}&l_2^2e^{2i\bar\beta}+
1&l_2^2e^{2i\bar\beta}\cr
                                    l_1l_2e^{i\bar\beta}&l_2^2e^{2i\bar\beta}
&l_2^2e^{2i\bar\beta}+1}.
\eqno(2.8)
$$
\noindent
We shall also employ the matrix 
$$
 h = m_D^\dagger m_D
\eqno(2.9)
$$
which is identical for the two textures of {\it Category} $A$ 
as well as for the two textures of {\it Category} $B$. 
Indeed, it can be given separately for the two categories as 
$$
h_A = \left|m_A\right|M_1\pmatrix{k_1^2&x^{1/4}k_1k_2e^{-i\bar\alpha}&
                                  x^{1/4}k_1k_2e^{-i\bar\alpha}\cr
             x^{1/4}k_1k_2e^{i\bar\alpha}&\sqrt{x}(1+k_2^2)&\sqrt{x}k_2^2\cr
             x^{1/4}k_1k_2e^{i\bar\alpha}&\sqrt{x}k_2^2&\sqrt{x}(1+k_2^2)},
\eqno(2.10a)
$$
$$
h_B = \left|m_B\right|M_1\pmatrix{l_1^2+2l_2^2&x^{1/4}l_2e^{-i\bar\beta}&
                                  x^{1/4}l_2e^{-i\bar\beta}\cr
                            x^{1/4}l_2e^{i\bar\beta}&\sqrt{x}&0\cr
                            x^{1/4}l_2e^{i\bar\beta}&0&\sqrt{x}},
\eqno(2.10b)
$$
where
$$
x = \frac{M_{2=3}^2}{M_1^2}.
\eqno(2.11)
$$ 
\n
Restrictions on the parameters $k_1$, $k_2$, $\cos\bar\alpha$ 
and $l_1$, $l_2$, $\cos\bar\beta$  
from neutrino oscillation data were worked 
out in ref. \cite{five}. The relevant measured quantities are the ratio 
of the solar to
atmospheric neutrino mass squared differences 
$R=\Delta m_{21}^2/\Delta m_{32}^2$ and the tangent of twice the solar mixing 
angle $\tan2\theta_{12}$. One can write 
$$
R = 2{(X_1^2+X_2^2)}^{1/2}{[X_3-{(X_1^2+X_2^2)}^{1/2}]}^{-1},
\eqno(2.12a)
$$
$$
\tan2\theta_{12} = \frac{X_1}{X_2}. 
\eqno(2.12b)
$$
\noindent
The quantities $X_{1,2,3}$ are given for the two categories as follows :
\vskip 0.1in
\noindent
{{\it Category}\,\, $A$} :
$$
X_{1A} = 2\sqrt{2}k_2{[{(1+2k_2^2)}^2 + k_1^4 + 2k_1^2(1+2k_2^2)\cos2\bar
\alpha]}^{1/2},
\eqno(2.13a)
$$
$$
X_{2A} = 1-k_1^4-4k_2^4-4k_1^2k_2^2\cos2\bar\alpha,
\eqno(2.13b)
$$
$$
X_{3A} = 1-4k_2^4-k_1^4-4k_1^2k_2^2\cos2\bar\alpha - 4k_2^2.
\eqno(2.13c)
$$
\vskip 0.1in
\noindent
{{\it Category}\,\,$B$} :
$$ 
X_{1B} = 2\sqrt{2}l_1l_2{[{(l_1^2+2l_2^2)}^2 + 
1+ 2(l_1^2+2l_2^2)\cos2\bar\beta]}^{1/2},
\eqno(2.13d)
$$
$$
X_{2B} = 1+4l_2^2\cos2\bar\beta+4l_2^4-l_1^4,
\eqno(2.13e)
$$
$$
X_{3B} = 1-{(l_1^2+2l_2^2)}^2 - 4l_2^2\cos2\bar\beta.
\eqno(2.13f)
$$
\noindent
We also choose to define 
$$
X_{A,B} = {({X_{1A,B}^2 + X_{2A,B}^2})}^{1/2}.
\eqno(2.14)
$$
\vskip 0.1in
\n
At the $3\sigma$ level, $\tan2\theta_{12}$ is presently known to be 
\cite{fourteen}  between 
1.83 and 4.90. For this range, only the inverted mass ordering for the light 
neutrinos, i.e. $\Delta m^2_{32}<0$, is allowed for {\it Category} 
$A$ with the allowed 
interval for $R$ being $-4.13\times {10}^{-2}\,\,eV^2$ 
to $-2.53\times {10}^{-2}\,\,eV^2$. In contrast, 
the same range of $\tan2\theta_{12}$ allows only the 
normal light neutrino mass ordering $\Delta m^2_{32}>0$ for {\it Category} $B$ 
with $R$
restricted  to be between $2.46\times {10}^{-2}\,\,eV^2$ and 
$3.92\times {10}^{-2}\,\,eV^2$. 
A thin sliver is allowed \cite{five} in the $k_1-k_2$ 
plane for {\it Category} $A$, while a substantial region with 
two branches is allowed \cite{five} in the $l_1-l_2$ plane for 
{\it Category} $B$. 
Finally, $\cos\bar\alpha$ is restricted to the interval bounded 
by $0$ and $0.0175$, while $\cos\bar\beta$ is restricted to 
the interval bounded by $0$ and $0.0523$. Thus, 
$\bar\alpha$, $\bar\beta$ could be either in the first or in the 
fourth quadrant. The interesting new point in the present 
work is that the baryogenesis constraint leads to restrictions on 
$\sin2\bar\alpha$ and $\sin2\bar\beta$ to the extent of removing the quadrant 
ambiguity in $\bar\alpha$ and $\bar\beta$.
\section{Basic calculation of baryon asymmetry}
\vskip 0.1in
\noindent
Armed with $\mu\tau$ symmetry as well as 
 eqs. (2.8) and (2.10), 
we can tackle leptogenesis at a scale $\sim$ $M_{lowest}$. 
There are three possible mass hierarchical 
cases for $N_i$. Case (a) corresponds to a normal hierarchy of the heavy 
Majorana neutrinos (NHN), i.e. $M_{lowest} = M_1<<M_2=M_3$. In case (b) one has an inverted hierarchy for $N_i$ (IHN) with $M_{lowest} = M_2=M_3<<M_1$. Case 
(c) refers to the quasidegenerate (QDN) situation with $M_1\sim M_2\sim M_3\sim M_{lowest}$. Working within the MSSM \cite{nine} and completely neglecting 
possible scattering processes \cite{ten} which violate lepton number, we can 
take the asymmtries generated by $N_i$ decaying into a doublet of leptons 
$L_{\alpha}$ and a Higgs doublet $H_u$ as 
\be
\epsilon^\alpha_i =\frac{\Gamma(N_i\rightarrow L_\alpha^C H_u)-
\Gamma(N_i\rightarrow L_\alpha H_u^C)}
{\Gamma(N_i\rightarrow L_\alpha^C H_u)+\Gamma(N_i\rightarrow L_\alpha H_u^C)}     \simeq
\frac{1}{4\pi v_u^2h_{ii}}\sum_{j\neq i}\left[{\cal I}_{ij}^\alpha f(x_{ij})+
{\cal J}_{ij}^\alpha \frac{1}{1-x_{ij}}\right],
\ee
\be
{\cal I}_{ij}^\alpha={\rm Im}\,\,
[{(m_D^\dagger)}_{i\alpha}{(m_D)}_{\alpha j}h_{ij}],
\ee
\be
{\cal J}_{ij}^\alpha={\rm Im}\,\,
[{(m_D^\dagger)}_{i\alpha}{(m_D)}_{\alpha j}h_{ji}],
\ee
\n
where 
\be
x_{ij} = M_{j}^2/M_i^2
\ee
\n
and 
\be
f(x_{ij}) = \sqrt{x_{ij}}\left[\frac{2}{1-x_{ij}} - 
{\rm ln}\frac{1+x_{ij}}{x_{ij}}\right ].
\ee
\vskip 0.1in
\noindent  
We note here that the ${\cal J}_{ij}^\alpha$ term does not
contribute to $\e_i^\alpha$ in our scheme since it
vanishes \cite{five} on account of $\mu\tau$ symmetry.
Further, contributions to $\epsilon_i^\alpha$ from $N_i$ decaying into 
sleptons and higgsinos and from 
sneutrinos $\tilde N_i$ decaying into sleptons 
and Higgs as well as into leptons and 
higgsinos have been included by appropriately 
choosing the $x_{ij}$-dependence in the RHS of eq. (3.5).
Observe also that ${\cal I}_{1j}^\alpha$ (and hence $\e_1^\alpha$)
gets an overall minus sign from Im$(e^{-i\bar\alpha}, e^{-i\bar\beta})$, 
whereas ${\cal I}_{2j}^{\alpha}$, ${\cal I}_{3j}^{\alpha}$ (and hence
$\e_{2,3}^\alpha$)
get an overall plus sign from Im$(e^{i\bar\alpha}, e^{i\bar\beta})$.
Except for being positive in the region $0.4\leq x_{ij}< 1$, the function 
$f(x_{ij})$ of eq.(3.5) is negative for all other values of its argument.
These signs are crucial in 
determining  the sign of $\eta$ and hence those of $\bar\alpha$, $\bar\beta$.
\vskip 0.1in
\noindent
The decay asymmetries $\epsilon_i^\alpha$ get converted into a 
lepton asymmetry 
$Y^\alpha = (n_l^\alpha-\bar n_l^\alpha)s^{-1}$, $s$ being the entropy density 
and $n_l^\alpha$ ($\bar n_l^\alpha$) 
being the leptonic (antileptonic) number density (including superpartners) 
for flavor $\alpha$  
via the washout relation \cite{ten} 
\be
Y^\alpha = \sum_i \epsilon_i^\alpha {\cal K}_i^\alpha g_{\star i}^{-1}.
\ee
\noindent
In eq. (3.6), $g_{\star i}$ is the effective number of spin degrees of freedom 
of particles and antiparticles 
at a temperature  equal to $M_i$. Furthermore, when all the flavors are active,
the quantity ${\cal K}_i^\alpha$ is given by the approximate 
relation \cite{add,thirteen}, neglecting contributions from off-diagonal 
elements of $A$,  
\be
{({\cal K}_i^\alpha)^{-1}}\simeq\frac{8.25}{|A^{\alpha\alpha}|K_i^\alpha} + 
{\left(\frac{|A^{\alpha\alpha}|K_i^\alpha}{0.2}\right)}^{1.16}.
\ee
In eq. (3.7), $K_i^\alpha$ is the flavor washout factor given by 
\be
K_i^\alpha=\frac{\Gamma\left(N\rightarrow L_\alpha H_u^C\right)}
{H(M_i)} =\frac{{|m_{D\alpha i}|}^2}{M_i}
\frac{M_{Pl}}{6.64\pi\sqrt{g_{\star i}}v_u^2},
\ee
\noindent
$M_{Pl}$ being the Planck mass. This follows 
since the Hubble expansion parameter $H(M_i)$ at a temperature $M_i$ is given 
by $1.66\sqrt{g_{\star i}} M_i^2 M_{Pl}^{-1}$.
Moreover, 
to the lowest order, 
$\Gamma(N_i \rightarrow L_\alpha H_u^C)$ equals  
${|m_{D\alpha i}|}^2 M_i {(4\pi v_u^2)}^{-1}$. 
An additional quantity, appearing in eq. (3.7), 
is $A^{\alpha\alpha}$, a diagonal element of the matrix 
$A^{\alpha\beta}$ defined by 
\be
Y_{\rm L}^\alpha = \sum_\beta A^{\alpha\beta}Y_\Delta^\beta.
\ee
\noindent
Here 
$Y_{\rm L}^\alpha = s^{-1}(n_{\rm L}^\alpha -{\bar{n}}_{\rm L}^\alpha)$, 
$n_L^\alpha$ being the number density of left-handed lepton and 
slepton doublets of flavor $\alpha$ and 
$Y_{\Delta}^\alpha = \frac{1}{3}Y_B - Y^\alpha$, $Y_B$ being the 
baryonic number density (normalized to the entropy density $s$) 
including all
superpartners. The precise forms for $A^{\alpha\beta}$ in 
different regimes of 
leptogenesis will be specified later.
\vskip0.1in
\noindent
One can now utilize the relation between $Y_B = {(n_B-n_{\bar B})}s^{-1}$ and 
$Y_l ={\sum_{\alpha}} Y^\alpha$, namely \cite{eleven} 
\be
Y_B = -\frac{8n_F+4n_H}{22n_F+13n_H}Y_l,
\ee
\noindent
where $n_F\,\, (n_H)$ is the number of matter fermion (Higgs) 
$SU(2)_L$ doublets present in the theory at electroweak temperatures. 
For MSSM, $n_F=3$ and $n_H=2$ so that eq. (3.10) becomes 
\be
Y_B = -\frac{8}{23}Y_l.
\ee
\n
The baryon asymmetry 
$\eta = (n_B-n_{\bar B})n_\gamma^{-1}$ can now be calculated, 
utilizing the result \cite{twelve} that $s n_\gamma^{-1}\simeq 7.04$ at the 
present time, to be 
\be
\eta = \frac{s}{n_\gamma}Y_B\simeq 7.04 Y_B\simeq -2.45 Y_l.
\ee
\n
Leptogenesis occurs at a temperature of the order of $M_{lowest}$ and 
the effective values of $A^{\alpha\alpha}$ and ${\cal K}_i^\alpha$ 
depend on which 
flavors are active in the washout process. 
This is controlled \cite{ten} by the quantity 
$M_{lowest}{(1+\tan^2\beta)}^{-1}$. There are three different regimes 
which we discuss separately.
\vskip 0.1in
\n 
{\bf{(1) ${\bf{M_{lowest}{(1+\tan^2\beta)}^{-1}>{10}^{12}}}$ GeV.}} 
\vskip 0.1in
\n
In this case there is no 
flavor discrimination and unflavored leptogenesis takes place. Thus 
$A^{\alpha\beta}=-\delta^{\alpha\beta}$ and all flavors ${\alpha}$ can just 
be summed in eqs. (3.1). Thus $\e_i = \sum\e_i^\alpha$, 
$\sum_\alpha{\cal J}_{ij}^\alpha = 0$, 
${{\cal I}_{ij}} \equiv \sum_\alpha {\cal I}_{ij}^\alpha = 
{\rm Im}\,\,{(h_{ij})}^2$ 
and $Y = 
\sum_i \epsilon_i{g_{\star i}^{-1}}{\cal K}_i$ with 
${\cal K}_i^{-1}=8.25 K_i^{-1}+{(K_i/0.2)}^{1.16}$ and 
$K_i =\sum_\alpha K_i^\alpha =  h_{ii} M_{Pl} {(6.64\pi \sqrt{g_{\star i}}M_iv_u^2)}^{-1}$. 
\n
For the normal hierarchical heavy neutrino (NHN) case (a), 
$M_{2=3}$ may be ignored and the index $i$ 
can be restricted to just 1, taking $g_{\star 1} = 232.5$. 
For the corresponding inverted hierarchical (IHN) case (b) 
$M_1$ can be ignored and $i$ made to run
over 2 and 3 with $g_{\star 2=3}$ = 236.25, all quantities involving the index 
2 being identical to the corresponding ones involving 3. Coming to the 
quasidegenerate (QDN) heavy neutrino case (c), $g_{\star}$ = 240 
and the contributions from $i=1$ must be separately 
added to identical contributions from 
$i=2,3$.
\vskip 0.1in
\n
{{\bf (2) ${\bf{M_{lowest}{(1+\tan^2\beta)}^{-1}<{10}^{9}}}$ GeV.}} 
\vskip 0.1in 
\n
Here, all flavors are 
separately active and one has fully flavored leptogenesis. Now the $A$-matrix
needs to be taken as \cite{ten} 
\be
A^{MSSM} = \pmatrix{-93/110 &  6/55  &   6/55\cr
                     3/40   & -19/30 &   1/30\cr
                     3/40   &  1/30  & -19/30}
\ee
\n
and eqs. (3.6) -- (3.8) used for each flavor $\alpha$. 
Once again, we consider the 
different cases (a), (b) and (c) of heavy neutrino mass ordering. Ignoring 
$M_{2=3}$ for case (a) and with $g_{\star 1} = 232.5$, we have $\eta\simeq 
-1.05\times {10}^{-2}\sum_{\alpha}\epsilon_1^\alpha{\cal K}_1^\alpha$. 
Similarly, ignoring $M_1$ for case (b) and with $g_{\star 2=3}=236.25$, 
one gets $\eta\simeq
-1.04\times {10}^{-2}\sum_{\alpha}(\epsilon_2^\alpha{\cal K}_2^\alpha+ 
\epsilon_3^\alpha{\cal K}_3^\alpha)$. For case (c), $g_{\star}=240$ and 
 $\eta\simeq
-1.02\times {10}^{-2}\sum_{\alpha}(\epsilon_1^\alpha{\cal K}_1^\alpha+      
\epsilon_2^\alpha{\cal K}_2^\alpha+
\epsilon_3^\alpha{\cal K}_3^\alpha)$.
\vskip 0.1in
\n
{\bf {(3) ${\bf{{10}^{9}\,\,{\rm{\bf{GeV}}}<M_{lowest}{(1+\tan^2\beta)}
^{-1}<{10}^{12}}}$ GeV.}} 
\vskip 0.1in
\n
In this regime the $\tau$- flavor decouples 
first while the electron and muon flavors act 
indistinguishably. The latter, therefore, can be summed. Now effectively $A$ 
becomes a $2\times 2$ matrix $\tilde A$ given by \cite{ten} 
\be 
\tilde A = \pmatrix{-541/761 &  152/761\cr
                     46/761  & -494/761}
\ee
\n
and acting in a space spanned by $e+\mu$ and $\tau$. Indeed, we can define 
${\cal K}_{i}^{e+\mu}$ and $\tilde{\cal K}_i^\tau$ by
$$ 
{({\cal K}_i^{e+\mu})^{-1}}=\frac{8.25}{|\tilde{A}^{11}|(K_i^e+K_i^\mu)} + 
{\left(\frac{|\tilde A^{11}|(K_i^e+K_i^\mu)}{0.2}\right)}^{1.16},
\eqno(3.15a)
$$
$$
{({\tilde{\cal K}}_i^{\tau})^{-1}}=\frac{8.25}{|\tilde{A}^{22}|K_i^\tau} + 
{\left(\frac{|\tilde A^{22}|(K_i^\tau)}{0.2}\right)}^{1.16}.
\eqno(3.15b)
$$
Now, for case (a) with $g_{\star 1}=232.5$, 
$\eta\simeq -1.05\times {10}^{-2}
[(\epsilon_1^e+\epsilon_1^\mu){\cal K}_1^{e+\mu}+\epsilon_1^\tau\tilde{\cal K}_1^\tau]$.
Case (b) has $g_{\star 2=3}$ = 236.25 and 
$\eta\simeq -1.04\times {10}^{-2}\sum_{k=2,3}
[(\epsilon_k^e+\epsilon_k^\mu){\cal K}_k^{e+\mu}
+\epsilon_k^\tau\tilde{\cal K}_k^\tau]$.
Finally, case (c), with  $g_{\star}$ = 240, 
has $\eta\simeq -1.02\times {10}^{-2}\sum_{i}
[(\epsilon_i^e+\epsilon_i^\mu){\cal K}_i^{e+\mu}
+\epsilon_i^\tau\tilde{\cal K}_i^\tau]$.
\section{Baryon asymmetry in the present scheme} 
\n
{\bf (1) Regime of unflavored leptogenesis}
\vskip 0.1in
\noindent
As explained in Sec. 3, there is no flavor discrimination if 
$M_{lowest}({1+\tan^2\beta)}^{-1}>10^{12}$ GeV. The lepton asymmetry parameters
$\epsilon_i$ can now be given after summing over $\alpha$. Additional simplifications can be made by taking $v_u=v\sin\beta$ with $v\simeq 246$ GeV and substituting 
\be
 |m| = {(\Delta m_{21}^2/X)}^{1/2}.
\ee
\n
The relevant expressions for the two categories then are the following
\vskip 0.1in
\n
{{\it Category} $A$} :
$$
\epsilon_{1A}\simeq -2.35\times {10}^{-8}
\frac{M_1}{{10}^9\,\,{\rm{GeV}}}{\frac 
{k_2^2\sqrt{x}f(x)\sin2\bar\alpha}{X_A^{1/2}\sin^2\beta 
}},
\eqno(4.2a)
$$
$$
\epsilon_{2A}=\epsilon_{3A}\simeq 1.18\times {10}^{-8}\frac{M_{2=3}}
{{10}^9\,\,{\rm{GeV}}}{\frac 
{k_1^2k_2^2\frac{1}{\sqrt{x}}f(\frac{1}{x})\sin2\bar\alpha}{(1+k_2^2)
{X_A^{1/2}\sin^2\beta} 
}}
$$
$$
= 1.18\times {10}^{-8}\frac{M_{1}}
{{10}^9\,\,{\rm{GeV}}}{\frac 
{k_1^2k_2^2 f(\frac{1}{x})\sin2\bar\alpha}{(1+k_2^2)
{X_A}^{1/2}\sin^2\beta}}. 
\eqno(4.2b)
$$
\n
{{\it Category} $B$}: 
$$
\epsilon_{1B}\simeq -2.35\times {10}^{-8}
\frac{M_1}{{10}^9\,\,{\rm{GeV}}}{\frac 
{l_2^2\sqrt{x}f(x)\sin2\bar\beta}{(l_1^2+2l_2^2)
X_B^{1/2}\sin^2\beta} 
},
\eqno(4.3a)
$$
$$
\epsilon_{2B}=
\epsilon_{3B}\simeq 1.18\times {10}^{-8}\frac{M_{2=3}}
{{10}^9\,\,{\rm{GeV}}}{\frac 
{l_2^2 f(\frac{1}{x})\sin2\bar\beta}{
X_B^{1/2}\sin^2\beta} 
}
$$
$$
= 1.18\times {10}^{-8}\frac{M_{1}}
{{10}^9\,\,{\rm{GeV}}}{\frac 
{l_2^2 \sqrt{x}f(\frac{1}{x})\sin2\bar\beta}{
X_B^{1/2}\sin^2\beta} 
}.
\eqno(4.3b)
$$
\n 
Note that $x$ was defined in eq.(2.11). We are now in a position to discuss the three $N_i$ mass hierarchical cases. 
For case (a), with the much heavier $M_2=M_3$ ignored and only 
$M_1$ contributing, we can give the following expressions  for 
the flavor-summed washout factors.   
\begin{flushleft}
{{\it Category} $A$} :
\end{flushleft}
$$ 
K_{1A}\simeq {\frac{86.36\,\, k_1^2}{{\sqrt{g_\star}}
X_A^{1/2}\sin^2\beta}},
\eqno(4.4a)
$$
$$
K_{2A}=K_{3A}\simeq{\frac{86.36\,\,(1+k_2^2)}{{\sqrt{g_\star}}
X_A^{1/2}\sin^2\beta}}.
\eqno(4.4b)
$$
\begin{flushleft} 
{{\it Category} $B$} : 
\end{flushleft}
$$
K_{1B}\simeq {\frac{86.36\,\, (l_1^2+2l_2^2)}
{{\sqrt{g_\star}}
X_B^{1/2}\sin^2\beta}},
\eqno(4.4c)
$$
$$
K_{2B} = K_{3B} \simeq\frac{86.36}{{\sqrt{g_\star}}
X_B^{1/2}\sin^2\beta} .
\eqno(4.4d)
$$
\n
Consequently,
$$
\eta^{NHN}_A \simeq -1.05\times {10}^{-2}\,\,{(\epsilon_{1A}{\cal K}_{1A})}
_{g_{\star=232.5}},
\eqno(4.5a)
$$ 
$$
\eta^{NHN}_B \simeq -1.05\times {10}^{-2}\,\,{(\epsilon_{1B}{\cal K}_{1B})}_{
g_{\star=232.5}},
\eqno(4.5b)
$$ 
\n
with the dependence on the category ($A$ or $B$) coming both through 
$\epsilon_1$ and $K_1$ occuring in ${\cal K}_1$. For case (b), 
one can ignore $M_1$ and hence
$\epsilon_1$ and ${\cal K}_1$. Thus we have 
$$
\eta^{IHN}_A \simeq -2.06\,\,{(\epsilon_{2A}{\cal K}_{2A})}_
{g_{\star=236.25}},
\eqno(4.6a)
$$ 
$$
\eta^{IHN}_B \simeq -2.06\,\,{(\epsilon_{2B}{\cal K}_{2B})}_
{g_{\star=236.25}},
\eqno(4.6b)
$$ 
\n
where once again the category dependence comes in through $\epsilon_2$ and 
$K_2$ occuring in ${\cal K}_2$. Finally, for case (c) with all three 
$M's$ contributing, 
$$
\eta_{A,B}^{QDN}\simeq -1.02\times {10}^{-2}\,\,{(\epsilon_{1A,B}{\cal K}_{1A,B}
+ 2\epsilon_{2A,B}{\cal K}_{2A,B})}_{g_{\star =240}}. 
\eqno(4.7)
$$
\n
The expressions for ${\cal K}_{1,2}$ in terms of $K_{1,2}$, have 
already been given in Sec. 3. Detailed expressions for the right hand 
sides of eqs. (4.5), (4.6) and (4.7) are given in appendix A.
\vskip 0.1in
\n
{\bf (2) Regime of fully flavored leptogenesis}
\vskip 0.1in
\n
If $M_{lowest}({1+\tan^2\beta)}^{-1}<10^{9}$ GeV, 
all leptonic flavors become active causing fully flavored leptogenesis, 
cf. Sec. 3. We now need to resort to eqs. (3.1) -- (3.8) to 
compute the lepton (flavor) asymmetry $Y^\alpha$. 
However, ${\cal J}_{ij}^\alpha$ vanishes explicitly for all the four cases 
of four zero  
textures of $m_D$ being considered by us. 
Thus we need be concerned only with the ${\cal I}_{ij}^\alpha$  term in 
eq. (3.1). Even some of the latter vanish on account of the zeroes 
in our textures. However, let us first draw some general 
conclusions about the two categories of textures before taking up the three 
$N_i$ hierarchical cases separately.
\vskip 0.1in
\n
{{\it Category} $A$}:
\vskip 0.1in
\n
It is clear from eq. (2.6a) that the presence of two zeroes in rows 2 and 
3 in both textures $m_{DA}^{(1)}$ and  $m_{DA}^{(2)}$ implies the vanishing of 
${(m_D)^\dagger}_{i\mu}{(m_D)}_{\mu j}$ and
 ${(m_D)^\dagger}_{i\tau}{(m_D)}_{\tau j}$ for $i\neq j$. 
As a result, $I_{ij}^\mu = I_{ij}^\tau = 0$ which imply that 
$$
\epsilon_{iA}^\mu = \epsilon_{iA}^\tau = 0 .
\eqno(4.8)
$$
\n
Thus $K_{iA}^\mu$ and $K_{iA}^\tau$ do not contribute to $\eta$. The expressions for the pertinent nonvanishing quantities are given by 
$$
\epsilon_{1A}^e\simeq -2.35\times {10}^{-8}
\frac{M_1}{{10}^9\,\,{\rm{GeV}}}\frac 
{k_2^2\sqrt{x}f(x)\sin2\bar\alpha}{X_A^{1/2}\sin^2\beta},
\eqno(4.9a)
$$
$$
\epsilon_{2A}^e=\epsilon_{3A}^e\simeq 1.18\times {10}^{-8}\frac{M_{1}}
{{10}^9\,\,{\rm{GeV}}}\frac 
{k_1^2k_2^2 f(\frac{1}{x})\sin2\bar\alpha}{(1+k_2^2)
X_A^{1/2}\sin^2\beta}\,\,,
\eqno(4.9b)
$$
$$
K_{1A}^e\simeq \frac{86.36\,\, k_1^2}{{\sqrt{g_\star}}
X_A^{1/2}\sin^2\beta}\,\, ,
\eqno(4.9c)
$$
$$
K_{2A}^e=K_{3A}^e\simeq\frac{86.36\,\,k_2^2}{{\sqrt{g_\star}}
X_A^{1/2}\sin^2\beta}\,\, .
\eqno(4.9d)
$$
\n
{{\it Category} $B$} :
\vskip 0.1in
\n
In this case, each allowed texture of $m_D$ in eq.(2.6b) has two zeroes in the first row in consequence of which 
${(m_D)^\dagger}_{ie}{(m_D)}_{ej}$ vanishes for $i\neq j$. Therefor, 
$I_{ij}^e =0$ because of which 
$$
\epsilon_{iB}^e = 0 .
\eqno(4.10)
$$
\n
in either case. Furthermore, with ${(h_B)}_{23}$ and ${(h_B)}_{32}$ being zero, $I_{23}^{\mu,\tau}$ and  $I_{32}^{\mu,\tau}$ vanish here for both textures $m_{DB}^{(1)}$ and 
 $m_{DB}^{(2)}$. An additional point is that, for the texture  $m_{DB}^{(1)}$,
$I_{12}^\mu = 0 = I_{13}^\tau$ but $I_{12}^\tau \neq 0 \neq I_{13}^\mu$ while,
for  $m_{DB}^{(2)}$, $I_{13}^\mu = 0 = I_{12}^\tau$ but  $I_{12}^\mu 
\neq 0 \neq I_{13}^\tau$. Consequently, $\epsilon_{1B}^\mu$ 
is the same for both allowed textures and so is  
$\epsilon_{1B}^\tau$. Moreover, for $m_{DB}^{(1)}$,
$\epsilon_{2B}^\mu$ and $\epsilon_{3B}^\tau$ vanish but $\epsilon_{2B}^\tau$ 
 and $\epsilon_{3B}^\mu$ do not while, for $m_{DB}^{(2)}$, 
$\epsilon_{2B}^\tau$ and $\epsilon_{3B}^\mu$ vanish but $\epsilon_{2B}^\mu$  
and $\epsilon_{3B}^\tau$ do not. In fact, explicitly one has
$$
\epsilon_{2B}^{(1)\mu}= \epsilon_{3B}^{(1)\tau}= \epsilon_{2B}^{(2)\tau}= 
\epsilon_{3B}^{(2)\mu} = 0,
\eqno(4.11a)
$$
$$
\epsilon_{1B}^{(1)\mu}= \epsilon_{1B}^{(1)\tau}= \epsilon_{1B}^{(2)\mu}= 
\epsilon_{1B}^{(2)\tau}\simeq 
-1.18\times {10}^{-8}\frac{M_{1}}
{{10}^9\,\,{\rm{GeV}}}\frac 
{l_2^2 \sqrt{x}f(x)\sin2\bar\beta}{(l_1^2+2l_2^2)
X_B^{1/2}\sin^2\beta} , 
\eqno(4.11b)
$$
$$
\epsilon_{2B}^{(1)\tau}= \epsilon_{3B}^{(1)\mu}= \epsilon_{2B}^{(2)\mu}= 
\epsilon_{3B}^{(2)\tau}\simeq 
1.18\times {10}^{-8}\frac{M_{2=3}}
{{10}^9{\rm{GeV}}}
\frac 
{l_2^2 {\frac{1}{\sqrt{x}}f(x)}\sin2\bar\beta}{
X_B^{1/2}\sin^2\beta}.
\eqno(4.11c)
$$
\n
In these equations and henceforth the superscripts 
(1),(2) refer to $m_D^{(1)}$, $m_D^{(2)}$ respectively. Coming to the washout factors, one sees a similar pattern. For $m_{DB}^{(1)}$, $K_{2B}^\mu$ and $K_{3B}^\tau$ vanish while for $m_{DB}^{(2)}$, $K_{2B}^\tau$ and $K_{3B}^\mu$ are zero.
Explicitly,
$$
K_{1B}^{(1)\mu}= K_{1B}^{(1)\tau}= K_{1B}^{(2)\mu} = K_{1B}^{(2)\tau}\simeq 
\frac{86.36\,\,l_2^2}{{\sqrt{g_\star}}
X_B^{1/2}\sin^2\beta} .
\eqno(4.12a)
$$  
$$
K_{2B}^{(1)e} = K_{2B}^{(2)e} = 
K_{2B}^{(1)\mu}= K_{3B}^{(1)\tau}= K_{2B}^{(2)\tau} =  
K_{3B}^{(2)\mu}= K_{3B}^{(1)e} = 
K_{3B}^{(2)e} = 0 ,
\eqno(4.12b)
$$
$$
K_{2B}^{(1)\tau}= K_{3B}^{(1)\mu}= K_{2B}^{(2)\mu} = K_{3B}^{(2)\tau}\simeq 
\frac{86.36}{{\sqrt{g_\star}}
X_B^{1/2}\sin^2\beta}.
\eqno(4.12c)
$$  
\n
Let us finally draw attention to an important consequence of 
eqs. (4.11) and (4.12). Since ${\cal K}_i^\alpha$ is just a 
known function of $A^{\alpha\alpha}$ as well as 
$K_i^\alpha$ and since $A^{\mu\mu}$ equals $A^{\tau\tau}$, the combination 
$$
\epsilon_{2B}^\mu{\cal K}_{2B}^\mu + \epsilon_{3B}^\mu{\cal K}_{3B}^\mu +
\epsilon_{2B}^\tau{\cal K}_{2B}^\tau + \epsilon_{3B}^\tau{\cal K}_{3B}^\tau
\eqno(4.13)
$$
\n
is identical for $m_{DB}^{(1)}$ and  
$m_{DB}^{(2)}$ and is a characteristic of just {\it Category} $B$.
\vskip 0.1in
\n
Now, for the normal $N_i$-hierarchical case (a), with $M_{2,3}$ neglected, 
we have the following expression for the baryon asymmetry. 
\begin{flushleft}
{{\it Category} $A$} : 
\end{flushleft}
$$
\eta_A^{NHN}\simeq -1.05\times {10}^{-2}\,\,\e_{1A}^e
{({\cal K}_{1A}^e)}_{g_{\star=232.5}}.
\eqno(4.14a)
$$
\begin{flushleft}
{{\it Category} $B$}: 
\end{flushleft}
$$
\eta_B^{NHN}\simeq  -1.05\times {10}^{-2}\,\,
{[(\e_{1B}^\mu{\cal K}_{1B}^\mu +\e_{1B}^\tau{\cal K}_{1B}^\tau)\\
 -2.1\times {10}^{-2}{(\e_{1B}^\mu{\cal K}_{1B}^\mu)}]}_{g_{\star=232.5}},
\eqno(4.14b)
$$
\n where $\mu\tau$ symmetry has been used in the last step. 
For the inverted $N_i$- hierarchical case (b), with $M_1$ neglected, 
the results are given below. 
\begin{flushleft}
{{\it Category} $A$} :
\end{flushleft}
$$
\eta_A^{IHN} \simeq 
-1.03\times {10}^{-2}\,\,{(\e_{2A}^e{\cal K}_{2A}^e + 
\e_{3A}^e{\cal K}_{3A}^e)}_{g_\star=236.25}
\simeq -2.06\times {10}^{-2} \e_{2A}^e {({\cal K}_{2A}^e)}_{g_\star=236.25}.
\eqno(4.15a)
$$
\begin{flushleft}
{{\it Category} $B$} :
\end{flushleft}
$$
\eta_B^{IHN} \simeq 
-1.03\times {10}^{-2}\,\,{(\e_{2B}^\mu{\cal K}_{2B}^\mu + 
\e_{2B}^\tau{\cal K}_{2B}^\tau + \e_{3B}^\mu{\cal K}_{3B}^\mu + 
\e_{3B}^\tau{\cal K}_{3B}^\tau)}_{g_\star=236.25}
$$$$
\simeq -2.06\times {10}^{-2}\,\,{(\e_{2B}^\mu {\cal K}_{2B}^\mu + 
\e_{3B}^\mu {\cal K}_{3B}^\mu)
}_{g_\star=236.25}.
\eqno(4.15b)
$$
\n
In eq. (4.15b), the first (second) term in the RHS bracket vanishes 
for $m_{DB}^{(1)}$  ($m_{DB}^{(2)}$); the nonvanishing terms have identical 
expressions for both textures. Lastly, for the quasidegenerate case (c), 
the expressions for the baryon asymmetry are as follows.
\begin{flushleft}
{{\it Category} $A$} :
\end{flushleft}
$$
\eta_A^{QDN} \simeq 
-1.02\times {10}^{-2}\,\,{(\e_{1A}^e{\cal K}_{1A}^e + 
2\e_{2A}^e{\cal K}_{2A}^e)}_{g_{\star = 240}}.
\eqno(4.16a)
$$
\begin{flushleft}
{{\it Category} $B$} :
\end{flushleft}
$$
\eta_B^{QDN} \simeq 
-1.02\times {10}^{-2}\,\,{(\e_{1B}^\mu{\cal K}_{1B}^\mu + 
\e_{1B}^\tau{\cal K}_{1B}^\tau + \e_{2B}^\mu{\cal K}_{2B}^\mu + 
\e_{2B}^\tau{\cal K}_{2B}^\tau + \e_{3B}^\mu{\cal K}_{3B}^\mu + 
\e_{3B}^\tau{\cal K}_{3B}^\tau)}_{g_\star=240}
$$
$$ 
\simeq -2.04\times {10}^{-2}\,\,{(\e_{1B}^\mu {\cal K}_{1B}^\mu + 
\e_{2B}^\mu {\cal K}_{2B}^\mu + \e_{3B}^\mu {\cal K}_{3B}^\mu)
}_{g_\star=240}.
\eqno(4.16b)
$$
\n
The second (third) term within the RHS bracket vanishes for $m_{DB}^{(1)}$ 
($m_{DB}^{(2)}$), while the remaining terms are identical for both textures of 
{\it Category} $B$. 
Detailed expressions for the right hand sides of eqs. (4.14), 
(4.15) and (4.16) appear in appendix B.
\vskip 0.1in
\noindent
{\bf (3) Regime of $\tau$-flavored leptogenesis}
\vskip 0.1in
\n
We have discussed in Sec. 3 that, with  
${10}^{9}\,\,{\rm{GeV}}<M_{lowest}{(1+\tan^2\beta)}^{-1}<{
10}^{12}\,\,{\rm GeV}$, there is flavor active 
leptogenesis in the $\tau$-sector but the electron 
and muon flavors can be summed. Thus, use can be made here of the flavor 
dependent results of Regime (2), but there is a proviso : 
both the generation and washout of $Y_L$ take place in a flavor 
subspace spanned by $e+\mu$ and $\tau$, 
cf. eqs. (3.13) and (3.14). Using the notation of eq. (3.15), 
we can then write the consequent baryon asymmetry as 
$$
\eta\simeq -2.45\sum_i g_{\star i}^{-1}[(\e_i^e+\e_i^\mu){\cal K}_i^{e+\mu}+
\e_i^\tau{\tilde{\cal K}_i^\tau}].
\eqno(4.17)
$$
\n
In discussing the lepton asymmetries and washout factors in detail here, 
it will 
be useful to consider the situation for each texture in either category by 
itself. We shall therefore separately enumerate the $N_i$-hierarchical cases 
(a), (b) and (c) for each of the four textures using the subscripts $A, B$ 
for the category and subscripts $(1),(2)$ for the textures. 
\vskip 0.1in
\n
{\it Category} $A$, $m_{DA}^{(1)}.$ 
\vskip 0.1in
\n
Now $\e_{iA}^{(1)\mu} = 0 = \e_{iA}^{(1)\tau}$, cf. eq. (4.8). But, 
in addition, we have
$$
0 = K_{1A}^{(1)\mu} =  K_{1A}^{(1)\tau} =  K_{2A}^{(1)\mu} =  K_{3A}^{(1)\tau}.
\eqno(4.18)
$$
\n
Here the nonvanishing $\e_{1A}^{(1)e}$,  $\e_{2A}^{(1)e}$ =  $\e_{3A}^{(1)e}$, 
$K_{1A}^{(1)e}$ and $K_{2A}^{(1)e} = K_{3A}^{(1)e}$ 
are as given by eqs. (4.9a) -- (4.9d). 
Additionally, 
$$
K_{1A}^{(1)e}=\frac{86.36}{\sqrt g_\star}
\frac{k_1^2}{X_A^{1/2}\sin^2\beta} ,
\eqno(4.19a)
$$
$$
K_{2A}^{(1)e} = K_{3A}^{(1)e} = \frac{86.36}{\sqrt g_\star}
\frac{k_2^2}{X_A^{1/2}\sin^2\beta} ,
\eqno(4.19b)
$$
$$
K_{3A}^{(1)\mu} = K_{2A}^{(1)\tau} = \frac{86.36}{\sqrt g_\star}
\frac{1}{X_A^{1/2}\sin^2\beta} .
\eqno(4.19c)
$$
\n Now for the NHN case (a), we have 
$$
\eta_A^{(1)NHN}\simeq -1.05\times {10}^{-2}
\e_{1A}^e
{\left[
{ \left({\cal K}_{1A}^{e+\mu}\right ) }_
{K_{1A}^\mu=0}
\right ]}
_{g_{\star=232.5}}
\eqno(4.20)
$$
\n with ${\cal K}_{1A}^{e+\mu}$ calculated as per eq. (3.15a) but setting 
$K_{1A}^\mu = 0$. For the IHN case (b), we can write 
$$
\eta_A^{(1)IHN}\simeq -1.03\times {10}^{-2}\,\,
{\left[
\e_{2A}^e
{ \left({\cal K}_{2A}^{e+\mu}\right ) }_
{K_{2A}^\mu=0} + \e_{3A}^e{\cal K}_{3A}^{e+\mu} 
\right ]}
_{g_{\star=236.5}} .
\eqno(4.21)
$$
\n Here again ${\cal K}_{2A}^{e+\mu}$ is calculated by putting 
$K_{2A}^\mu$ = 0. 
\vskip 0.1in
\n For the QDN case (c), the expression is 
$$
\eta_A^{(1)QDN}\simeq -1.02\times {10}^{-2}\,\,
{\left[
\e_{1A}^e
{ \left({\cal K}_{1A}^{e+\mu}\right ) }_
{K_{1A}^\mu=0} 
+ \e_{2A}^e{\left({\cal K}_{2A}^{e+\mu}\right)}_{K_{2A}^\mu=0}
+ \e_{3A}^e{{\cal K}_{3A}^{e+\mu}}
\right ]}
_{g_{\star=240}}.
\eqno(4.22)
$$
\n
Once more, appropriate washout factors have to be set at zero as shown earlier 
in the calculation of  ${\cal K}_{iA}^{e+\mu}$. 
\vskip 0.1in
\n
{\it Category}\,\, $A$,\,\, $m_{DA}^{(2)}$.
\vskip 0.1in
\n
Again, $\e_{iA}^{(2)\mu} = 0 = \e_{iA}^{(2)\tau}$, but the vanishing 
washout factors now are 
$$
0 = K_{1A}^{(2)\mu} =  K_{1A}^{(2)\tau} =  K_{3A}^{(2)\mu} =  K_{2A}^{(2)\tau}.
\eqno(4.23)
$$
\n The pertinent nonzero quantities namely,  $\e_{1A}^{(2)e}$, 
$\e_{2A}^{(2)e} = \e_{3A}^{(2)e}$, $K_{1A}^{(2)e}$, 
$K_{2A}^{(2)e}$ and $K_{3A}^{(2)e}$ are the same as for $m_{DA}^{(1)}$. 
In addition, 
$$
K_{2A}^{(2)\mu} = K_{3A}^{(2)\tau} = \frac{86.36}{\sqrt g_\star}
\frac{1}{X_A^{1/2}\sin^2\beta}.
\eqno(4.24)
$$
\n Thus, for the NHN case (a), 
$$
\eta_A^{(2)NHN}\simeq -1.05\times {10}^{-2}\,\,
\e_{1A}^e
{\left[
{\left({\cal K}_{1A}^{e+\mu}\right)}_
{K_{1A}^\mu=0}
\right]}
_{g_{\star=232.5}},
\eqno(4.25)
$$
\n i.e. the same as in eq. (4.20). Then, for the IHN case (b), we have 
$$
\eta_A^{(2)IHN}\simeq 1.03\times {10}^{-2}\,\,
{\left[
\e_{2A}^e
	{{\cal K}_{2A}^{e+\mu}}+ \e_{3A}^e{\left({\cal K}_{3A}^{e+\mu}\right)}_
{K_{3A}^\mu=0} 
\right ]}
_{g_{\star=236.25}} ,
\eqno(4.26)
$$
\n i.e. ${\cal K}_{2A}^{e+\mu}$ is calculated fully 
but  ${\cal K}_{3A}^{e+\mu}$
by setting $K_{3A}^\mu$ = 0. This expression turns out to be the same as for 
$m_{{DA}}^{(1)}$. 
\vskip 0.1in  
\n Finally, the QDN case (c) has  the baryon asymmetry as 
$$
\eta_A^{(2)QDN}\simeq -1.02\times {10}^{-2}\,\,
{\left[
\e_{1A}^e
{\left({\cal K}_{1A}^{e+\mu}\right)}_
{K_{1A}^\mu=0} + \e_{2A}^e{\cal K}_{2A}^{e+\mu}
+ \e_{3A}^e{\left({\cal K}_{3A}^{e+\mu}\right)}_{K_{3A}^\mu=0}
\right]}
_{g_{\star=240}}
\eqno(4.27)
$$
\n with appropriate washout factors set to zero in ${\cal K}_{iA}^{e+\mu}$, 
as shown. Again, this turns out to be equal to that for 
$m_{DA}^{(1)}$. Detailed expressions for the right hand side of eqs. (4.20) 
and (4.25), which are identical, appear in appendix C. 
We make the same statement for eqs. (4.21) and (4.26) as well as for 
eqs. (4.22) and (4.27). 
\vskip 0.1in
\n
{{\it Category} $B$, $m_{DB}^{(1)}$.}
\vskip 0.1in
\n
Here, $\e_{iB}^{(1)e}=0=\e_{2B}^{(1)\mu} = \e_{3B}^{(1)\tau}$ and 
$K_{2B}^{(1)e}$ cf. eqs. (4.10) 
and (4.11a), while the vanishing washout factors are $K_{2B}^{(1)\mu}$, 
$K_{3B}^{(1)\tau}$, $K_{3B}^{(1)e}$. The pertinent nonzero 
quantities, as given in eqs. (4.11b), (4.11c) and (4.12a), (4.12c), are 
$\e_{1B}^{(1)\mu}$ = $\e_{1B}^{(1)\tau}$, 
$\e_{2B}^{(1)\tau}$ = $\e_{3B}^{(1)\mu}$ and 
$K_{1B}^{(1)\mu}$ = $K_{1B}^{(1)\tau}$, 
$K_{2B}^{(1)\tau}$ = $K_{3B}^{(1)\mu}$. Additionally, 
$$
K_{1B}^{(1)e} = \frac{86.36}{\sqrt g_\star}
\frac{l_1^2}{X_B^{1/2}\sin^2\beta} ,
\eqno(4.28a)
$$
$$
K_{1B}^{(1)\mu} = K_{1B}^{(1)\tau} = \frac{86.36}{\sqrt g_\star}
\frac{l_2^2}{X_B^{1/2}\sin^2\beta} .
\eqno(4.28b)
$$
$$
K_{2B}^{(1)\tau} = K_{3B}^{(1)\mu} = \frac{86.36}{\sqrt g_\star}
\frac{1}{X_B^{1/2}\sin^2\beta} .
\eqno(4.28c)
$$
\n Therefore, for the NHN case (a), 
$$
\eta_B^{(1)NHN}\simeq - 1.05\times {10}^{-2}\,\,
{\left[
\e_{1B}^\mu
{\cal K}_{1B}^{e+\mu}+ 
\e_{1B}^\tau{\tilde{\cal K}}_{1B}^\tau 
\right]}
_{g_{\star=232.5}}.
\eqno(4.29)
$$
\n 
Coming to the IHN case (b), we have 
$$
\eta_B^{(1)IHN}\simeq -1.03\times {10}^{-2}\,\,
{\left[
\e_{3B}^\mu
{\left({\cal K}_{3B}^{e+\mu}\right)}_
{K_{3B}^{\mu}=0} + \e_{2B}^\tau{{\tilde{\cal K}}}_{2B}^{\tau} 
\right]}
_{g_{\star=236.25}}, 
\eqno(4.30)
$$
\n 
with the first term within the RHS bracket calculated by setting 
$K_{3B}^\mu =0$. For the final QDN case (c), the expression is 
$$
\eta_B^{(1)QDN}\simeq -1.02\times {10}^{-2}\,\,
{\left[
\e_{1B}^\mu{\cal K}_{1B}^{e+\mu} + \e_{3B}^\mu
{ \left({\cal K}_{3B}^{e+\mu}\right) }_
{K_{3B}^e=0} + \e_{1B}^\tau{\tilde{\cal K}}_{1B}^{\tau} + 
\e_{2B}^\tau{\tilde{\cal K}}_{2B}^{\tau} 
\right]}
_{g_{\star=240}},
\eqno(4.31)
$$
where ${\cal K}^{e+\mu}_{3B}$ is calculated with $K_{3B}^e$ set to vanish.
\vskip 0.1in
\n
{{\it Category} $B$, $m_{DB}^{(2)}$.}
\vskip 0.1in
\n
Here we have $\e_{iB}^{(2)e} = 0 = \e_{2B}^{(2)\tau} = \e_{3B}^{(2)\mu}$ 
from eqs. (4.10) and (4.11a), while the
washout factors 
$
K_{2B}^{(2)\tau}$, $K_{3B}^{(2)\mu}$,  $K_{2B}^{(2)e}$, $K_{3B}^{(2)e}$ 
vanish.
The remaining nonzero quantities of relevance, as appear in eqs. (4.11b), (
4.11,c) and 
(4.12a) (4.12c), are   
$\e_{1B}^{(2)\mu}$
= $\e_{1B}^{(2)\tau}$, $\e_{2B}^{(2)\mu}$ = $\e_{3B}^{(2)\tau}$ and 
$K_{1B}^{(2)\mu} = K_{1B}^{(2)\tau}$, $K_{3B}^{(2)\tau}=K_{2B}^{(2)\mu}$.
In addition, 
$K_{1B}^{(2)e}$ has the same expression as $K_{1B}^{(1)e}$ i.e.  
$$
K_{1B}^{(2)e} = \frac{86.36}{\sqrt g_\star}
\frac{l_1^2}{X_B^{1/2}\sin^2\beta} .
\eqno(4.32)
$$
\n The NHN case (a) now yields  
$$
\eta_B^{(2)NHN}\simeq -1.05\times {10}^{-2}
{\left[\e_{1B}^\mu{\cal K}_{1B}^{e+\mu} + 
\e_{1B}^\tau{\tilde{\cal K}}_{1B}^{\tau}
\right]}_{g_\star = 232.5},
\eqno(4.33)
$$
\n 
as with $m_{DB}^{(1)}$. For the IHN case (b), the baryon asymmetry reads 
$$
\eta_B^{(2)IHN}\simeq -1.03\times {10}^{-2}\,\,
{\left[
\e_{3B}^\tau
{ \left({\tilde{\cal K}}_{3B}^{\tau}\right ) 
+ \e_{2B}^\mu\left({{\cal K}_{2B}^{e+\mu}}\right)}_
{K_{2B}^e=0} 
\right ]}_{g\star = 236.25}
\eqno(4.34)
$$
\n 
which happens to have  the same expression as for $m_{DB}^{(1)}$. Finally, 
for the QDN case (c), the baryon asymmetry is 
$$
\eta_B^{(2)QDN}\simeq -1.02\times {10}^{-2}\,\,
{\left[
\e_{1B}^\mu{\cal K}_{1B}^{e+\mu} + \e_{2B}^\mu
{ \left({\cal K}_{2B}^{e+\mu}\right) }_
{K_{2B}^{e}=0} + \e_{1B}^\tau{\tilde{\cal K}}_{1B}^{\tau} + 
\e_{3B}^\tau{\tilde{\cal K}}_{3B}^{\tau} 
\right]}_{g\star = 240}
\eqno(4.35)
$$
which also turns out to be the same as for 
$m_{DB}^{(1)}$. Thus the baryon asymmetry in each of the 
three $N_i$-hierarchical cases has the same expression for 
both $m_D^{(1)}$ and $m_D^{(2)}$ in {\it Category} $A$ and the same statement 
holds for {\it Category} $B$. 
Detailed expressions of $\eta$ in {\it Category} $B$ for the NHN, 
IHN and QDN cases 
are given in appendix C.
\section{Results and discussion}
\n
We had earlier deduced
\cite{five} from neutrino oscillation data with 
$3\sigma$ errors the constraints 
$0\leq \cos\bar\alpha\leq 0.0175$ and $0\leq \cos\bar\beta\leq 0.0523$ for the 
phases $\bar\alpha$ and $\bar\beta$ of {\it Categories} $A$ and $B$ 
respectively. 
Thus each phase could have been in either the first or the fourth quadrant with 
$89^o\leq |\bar\alpha|\leq 90^o$ and $87^o\leq |\bar\beta|\leq 90^o$.
The new requirement of matching the generated baryon asymmetry $\eta_A$ 
($\eta_B$) for {\it Category} $A$ ($B$) with its observed value in the  
 $3\sigma$ range
$5.5\times {10}^{-10}$ to $7.0\times {10}^{-10}$ 
\cite{eleven}--\cite{trodden} 
puts restrictions on 
$\sin2\bar\alpha$ $(\sin2\bar\beta)$ which fix both the magnitude 
and the sign of 
$\bar\alpha$ $(\bar\beta)$. To be specific in our 
numerical analysis, we choose $x=M_{2=3}^2/M_{1}^2$ for the 
different hierarchical cases as follows :
(a) for NHN, $x\ge 10$, (b) for IHN, $x\le 0.1$, 
(c) for QDN, $0.1\le x\le 10$. So far, we did not dwell on the 
mass ordering (normal or inverted)
 of the right handed heavy neutrinos $N_i$ in the QDN case. 
For the normal ordering (NON) case, we take 
$1.1\le x\le 10$, while for an inverted ordering (ION), 
our choice is $0.1\le x\le 0.9$. 
As mentioned earlier, the function 
$f(x)$ is 
positive for $0.4\leq x< 1.0$ and negative elsewhere.
We need to avoid the point $x=1$ which corresponds 
to the complete degeneracy of the $N_i$, i.e. 
$M_1=M_2=M_3$ since $f(x)$ diverges at this point. The inclusion of  
finite width corrections  to propagators of right handed 
neutrinos in the one loop decay diagrams avoids 
this problem. Now, both the previously 
divergent part of the modified $f(x)$ and
 the lepton asymmetry vanish there. We also avoid the near $x=1$ region, 
$0.9< x< 1.1$, to exclude the 
so called resonant leptogenesis \cite{resolepto} since that is not part 
of our scenario.   
\noindent
Tables 1 -- 3 enumerate the emergent constraints 
on $\bar\alpha$, $\bar\beta$ in consequence of matching $\eta_A$, $\eta_B$ 
\begin{table}[htb]   
\begin{center}  
{\small
\begin{tabular}{|c|c|c|c|c|}
\hline
\multicolumn{5}{|c|}{Category $A$} \\
\hline
Parameters&NHN&IHN&\multicolumn{2}{|c|}{QDN}\\\cline{4-5}
&&&NON&ION\\
\hline
&&&&\\
$\bar\alpha$&$\bar\alpha<0$&$\bar\alpha>0$&$\bar\alpha<0$
&$\bar\alpha>0$\\
&$89.0^o-89.9^o$&$89.95^o-89.99^o$& 
$89.1^o-89.9^o$&$89.10^o-89.99^o$\\
\hline
$x$& $10-10^3$ & $0.001-0.1$  
& $2.0 - 9.1$ & $0.1-0.9$ \\
\hline
$\tan\beta$& $2-60$&$2-5$&$2-60$&$2-60$\\
\hline
&$5.0\times {10}^{3}$&$5.0\times{10}^3$&
$5\times {10}^3$&$5.0\times{10}^3$\\
$\frac{M_{lowest}}{{10}^9GeV}$&---&---&---&--- \\
& $4.9\times{10}^6$ & $2.6\times{10}^4$ & $3.6\times{10}^6$ &
$4.9\times{10}^6$\\
\hline
\multicolumn{5}{|c|}{Category $B$} \\
\hline
Parameters&NHN&IHN&\multicolumn{2}{|c|}{QDN}\\\cline{4-5}
&&&NON&ION\\
\hline
&&&&\\
$\bar\beta$&$\bar\beta<0$&$\bar\beta>0$&$\bar\beta<0$
&$\bar\beta>0$\\
&$88.8^o-89.9^o$&$89.48^o-89.99^o$& 
$87.0^o-89.9^o$&$89.84^o-89.99^o$\\
\hline
$x$& $10-10^3$ & $0.001-0.1$  
& 8.3 - 9.5 & $0.1-0.9$ \\
\hline
$\tan\beta$& $2-8$&$2-12$&$2-60$&$2-10$\\
\hline
&$8.4\times {10}^{3}$&$5.0\times{10}^3$&
$5.0\times {10}^3$&$5.0\times {10}^3$\\
$\frac{M_{lowest}}{{10}^9GeV}$&---&---&---&--- \\
& $8.5\times{10}^4$ & $1.6\times{10}^5$ & $4.9\times{10}^6$ &
$1.0\times{10}^5$\\
\hline
\end{tabular}
\caption{Allowed $\bar\alpha$,$\bar\beta$ and 
other parameters for unflavored leptogenesis}
}
\end{center}
\end{table}
\begin{table}[htb]   
\begin{center}  
{\small
\begin{tabular}{|c|c|c|c|c|}
\hline
\multicolumn{5}{|c|}{Category $A$} \\
\hline
Parameters&NHN&IHN&\multicolumn{2}{|c|}{QDN}\\\cline{4-5}
&&&NON&ION\\
\hline
&&&&\\
$\bar\alpha$&$\bar\alpha<0$&$\bar\alpha>0$&$\bar\alpha<0$
&$\bar\alpha>0$\\
&$89.4^o-89.9^o$&$89.0^o-89.8^o$& 
$89.1^o-89.9^o$&$89.0^o-89.9^o$\\
\hline
$x$& $10-10^3$ & $0.001-0.1$  
& $1.1 - 10.0$ & $0.1-0.9$ \\
\hline
$\tan\beta$& $25-60$&$22-60$&$2-60$&$2-60$\\
\hline
&$67$&$4.9\times{10}^2$&
$23$&$10$\\
$\frac{M_{lowest}}{{10}^9GeV}$&---&---&---&--- \\
& $3.6\times{10}^3$ & $3.6\times{10}^3$ & $3.60\times{10}^3$ &
$3.6\times{10}^3$\\
\hline
\multicolumn{5}{|c|}{Category $B$} \\
\hline
Parameters&NHN&IHN&\multicolumn{2}{|c|}{QDN}\\\cline{4-5}
&&&NON&ION\\
\hline
&&&&\\
$\bar\beta$&$\bar\beta<0$&$\bar\beta>0$&$\bar\beta<0$
&$\bar\beta>0$\\
&$87.0^o-89.9^o$&$87.0^o-89.9^o$& 
$87.0^o-89.9^o$&$87.0^o-89.9^o$\\
\hline
$x$& $10-10^3$ & $0.001-0.1$  
& $1.1 - 10$ & $0.3-0.9$ \\
\hline
$\tan\beta$& $16-60$&$24-60$&$6-60$&$7-60$\\
\hline
&$2.4\times {10}^{2}$&
$5.7\times {10}^2$&$0.35\times {10}^{2}$& $0.49\times {10}^2$\\
$\frac{M_{lowest}}{{10}^9GeV}$&---&---&---&--- \\
& $3.6\times{10}^3$ & $3.6\times{10}^3$ & $3.6\times{10}^3$ &
$3.6\times{10}^3$\\
\hline
\end{tabular}
\caption{Allowed $\bar\alpha$, $\bar\beta$ and other parameters for fully 
flavored leptogenesis}
}
\end{center}
\end{table}
\begin{table}[htb]   
\begin{center}  
{\small
\begin{tabular}{|c|c|c|c|c|}
\hline
\multicolumn{5}{|c|}{Category $A$} \\
\hline
Parameters&NHN&IHN&\multicolumn{2}{|c|}{QDN}\\\cline{4-5}
&&&NON&ION\\
\hline
&&&&\\
$\bar\alpha$&$\bar\alpha<0$&$\bar\alpha>0$&$\bar\alpha<0$
&$\bar\alpha>0$\\
&$89.0^o-89.9^o$&$89.0^o-89.9^o$& 
$89.0^o-89.9^o$&$89.0^o-89.9^o$\\
\hline
$x$& $10-10^3$ & $0.001-0.1$  
& $1.1 - 10.0$ & $0.1-0.9$ \\
\hline
$\tan\beta$& $2-60$&$2-60$&$2-60$&$2-60$\\
\hline
&$1.7\times 10^3$&$50$&
$100$&$100$\\
$\frac{M_{lowest}}{{10}^9GeV}$&---&---&---&--- \\
& $4.0\times{10}^4$ & $1.03\times{10}^4$ & $1.97\times{10}^4$ &
$1.45\times{10}^4$\\
\hline
\multicolumn{5}{|c|}{Category $B$} \\
\hline
Parameters&NHN&IHN&\multicolumn{2}{|c|}{QDN}\\\cline{4-5}
&&&NON&ION\\
\hline
&&&&\\
$\bar\beta$&$\bar\beta<0$&$\bar\beta>0$&$\bar\beta<0$
&$\bar\beta>0$\\
&$87.0^o-89.9^o$&$87.0^o-89.9^o$& 
$87.0^o-89.9^o$&$87.0^o-89.9^o$\\
\hline
$x$& $10-10^3$ & $0.001-0.1$  
& $1.1 - 10.0$ & $0.1-0.9$ \\
\hline
$\tan\beta$& $2-60$&$2-60$&$2-60$&$2-60$\\
\hline
&$3.25\times 10^2$&
$6.25\times {10}^2$&$0.37\times {10}^{2}$& $0.37\times {10}^{2}$\\
$\frac{M_{lowest}}{{10}^9GeV}$&---&---&---&--- \\
& $2.3\times{10}^4$ & $5.0\times{10}^4$ & $2.1\times{10}^4$ &
$1.6\times{10}^5$\\
\hline
\end{tabular}
\caption{Allowed $\bar\alpha$, 
$\bar\beta$ and other parameters for $\tau$-flavored leptogenesis}
}
\end{center}
\end{table}
for each of the eighteen different possibilities described earlier 
with corresponding 
restrictions on the parameters $x$, $\tan\beta$ and $M_{lowest}$ as shown.
We would like to make the following 
comments on the information contained in tables 
1 -- 3.
\vskip 0.1in
\n
\begin{enumerate}

\item {\it Signs of phase angles}\, : We have a positive baryon asymmetry 
in our universe. From the formulae for
 all NHN cases in the Apendices, 
we can say that sign of $f(x)\sin2({\bar\alpha,\bar\beta})$ 
has to be positive in order to generate  such  
a positive asymmetry. But $f(x)$ is negative in the 
NHN region of $x\ge 10$. So, $\bar\alpha,~\bar\beta$ have to be 
negative for all NHN
cases. On the contrary, for all $IHN$ cases, there is an overall negative 
sign in the formulae for $\eta$ since 
${\rm Im}(h^2_{21})={\rm Im}(h^2_{31})$ here is opposite in sign to 
${\rm Im}(h^2_{12})={\rm Im}(h^2_{13})$ that come in for the NHN case. 
So, for a positive $\eta$, a 
negative sign of $f(x)\sin2({\bar\alpha,\bar\beta})$ is needed in 
 all IHN cases. Again,
$f(x)$ is negative in the NHN region of $x\le 0.1$. For this reason, 
 $\bar\alpha,~\bar\beta$ are positive in all IHN cases. 
For QDN cases we need to discuss 
the possibilities of normal and inverted ordering of $M_i$ separately. 
Here there are two terms with $f(x)$ and $-f(1/x)$ along with an overall 
factor $\sin2({\bar\alpha,\bar\beta})$. For the NON 
region $1.1\le x\le 10.0$, $f(x)$ is negative while 
$-f(1/x)$ is negative for $1.1\le x\leq 2.5$.
So, for the region $1.1\le x\leq 2.5$, 
$\bar\alpha,~\bar\beta$ are required to be negative. For the remaining 
part of the NON  region $2.5\leq x\le 10$, $f(x)$ 
is negative and $-f(1/x)$ positive but the $f(x)$ term 
dominates over the 
$-f(1/x)$ term. So, negative signs also are 
needed for $\bar\alpha,~\bar\beta$, 
in the region  $2.5\leq x\le 10$. Thus all QDN cases with NON 
require negative sign of $\bar\alpha,~\bar\beta$. Again, for QDN with
ION, both $f(x)$ and $-f(1/x)$ are positive 
in  the region $0.4\leq x\le 0.9$. 
In the rest of the ION region $0.1\le x\leq 0.4$, $f(x)$ is negative
and $-f(1/x)$ is positive. But, now the latter term dominates over the former 
one. So, positive  $\bar\alpha,~\bar\beta$ are needed in all QDN
cases with ION. In fact, we see  (tables 1 -- 3) that for all normal 
(both hierarchical and quasidegenrate) mass ordering 
 cases of $M_i$, the phases are negative  whereas,
for all inverted (both hierarchical and quasidegenrate) mass ordering cases, 
they are positive. One may also note that in all cases and regimes the 
size of the allowed range of $\tan\beta$ is correlated with that of the phase 
$\bar\alpha$/$\bar\beta$.

\item {\it Magnitudes of phase angles and other parameters}\, : 
Neither $\bar\alpha$ nor $\bar\beta$ can be strictly $90^o$ since 
$\eta$ then vanishes. Therefore, a nonzero $\eta$ is incompatible 
in {\it Category} $A$ with tribimaximal mixing which requires \cite{five}
 $\bar\alpha\,\, =\,\, \pi/2$. The numerical 
value of $\eta$ is most sensitive to the values of 
$\sin2(\bar\alpha,\bar\beta)$, $M_{lowest}$ ($M_{1}$ for normal mass ordering,
$M_2$ for inverted mass ordering) and to 
some extent to the function $f$ (and hence $x$) for acceptable ranges of 
$k_1,~k_2$ ({\it Category} A) and $l_1,~l_2$ (
{\it Category B}). The latter are of course restricted \cite{five} 
by the neutrino oscillation data.
For unflavored leptogenesis with 
$M_{lowest}>(1+\tan^2\beta)10^{12}$ GeV, the minimum value of 
$M_{lowest}$ is $5\times 10^{12}$ GeV, while we cut the maximum 
value at $5\times 10^{15}$ GeV to avoid the 
GUT scale whereabouts all produced asymmetry 
gets washed out by inflation. 
Such a large value of $M_{lowest}$ forces a small value of 
$\sin2(\bar\alpha,\bar\beta)$ in order 
to have the baryon asymmetry in the right range.
In {\it Category} $A$, the range of $|\bar\alpha|$ is restricted to 
$89^o\leq |\bar\alpha|\leq 90^o$ so that  $\sin2\bar\alpha$ is small there.
In the IHN case of {\it Category} $A$, other associated factors 
including $f(1/x)$ cause 
further restrictions on $\bar\alpha$, cf. Table 1. 
In {\it Category} $B$  the 
range $87^o\leq |\bar\beta|\leq 90^o$ is curtailed to 
$|\bar\beta|>88.8^o$ due to the large value of $M_{lowest}$ 
in flavor independent leptogenesis except the QDN (NON) case where other 
factors are responsible for necessary suppression.  
\vskip 0.1in
\n
\item  The quadrants 
of $\bar\alpha$, $\bar\beta$ do not change between 
unflavored, fully flavored and $\tau$-flavored leptogenesis, nor is there 
any dependence of them on the value of $\tan\beta$. 
They only depend on whether $N_i$ have a 
normal $(M_1<M_{2=3})$ or inverted $(M_1>M_{2=3})$ mass ordering. 
For the former, $\bar\alpha$ and 
$\bar\beta$ are always in the fourth quadrant $(<0)$ since $\e_1$ always 
has a minus sign in front, while 
the latter always forces them to be in the first quadrant $(>0)$ since 
$\e_2$ = $\e_3$ always has a plus sign in front. 
\vskip 0.1in
\n
\item
The 
constraints on $\sin2\bar\alpha$, $\sin2\bar\beta$ - extracted from 
$\eta_{A,B}$ - 
restrict the allowed intervals for $|\bar\alpha|$, $|\bar\beta|$ more 
stringently than do constraints on $\cos\bar\alpha$,  
$\cos\bar\beta$ obtained \cite{five}
from neutrino oscillation phenomenology.
\end{enumerate}
\section{Effect of radiative $\mu\tau$ symmetry breaking}
\n
While explaining a maximal value for $\theta_{23}$, exact $\mu\tau$ symmetry 
predicts a vanishing $\theta_{13}$. The latter will make the CP violating 
Dirac phase $\delta_D$ unobservable in neutrino oscillation experiments, 
many of which are being planned to study CP violation in the neutrino sector. 
Thus it may be desirable to have a nonzero $\theta_{13}$, however small. 
\vskip 0.1in
\n
Suppose $\mu\tau$ symmetry is exact at a high energy $\Lambda\sim {10}^{12}$ 
GeV characterizing the heavy Majorana neutrino mass scale. Running down to 
a laboratory scale $\lambda\sim {10}^3$ GeV, via one-loop renormalization 
group evolution, one picks up small factorizable departures from 
$\mu\tau$ symmetry, induced by charged lepton mass terms, in the 
elements of the light neutrino mass matrix $m_\nu$. These cause 
small departures 
from $45^o$ in $\theta_{23}^\lambda$ and tiny nonzero values for 
 $\theta_{13}^\lambda$. Neglecting $m_{\mu,e}^2$ in comparison with 
$m_\tau^2$, one obtains \cite{five} that
$$
m_\nu^\lambda \simeq \pmatrix{1&0&0\cr
                              0&1&0\cr 
                              0&0&1 - \Delta_\tau} 
m_\nu^\Lambda\pmatrix{1&0&0\cr
          0&1&0\cr
          0&0&1 - \Delta_\tau},
\eqno(6.1)
$$
\n
where $m_\nu^\Lambda$ is $\mu\tau$ symmetric and 
the deviation $\Delta_\tau$ is given in MSSM by 
$$
\Delta_\tau \simeq 
\frac{m_\tau^2}{8\pi^2v^2}(1+\tan^2\beta){\rm ln}\frac{\Lambda}{\lambda} 
\simeq 
6\times {10}^{-6}\,\,(1+\tan^2\beta).
\eqno(6.2)
$$ 
\n Working to the lowest nontrivial order in $\Delta_\tau$, the 
phenomenological consequences of eq. (6.1), derived from 
extant neutrino oscillation data, 
were worked out in ref.[15]. The 
allowed regions in the $k_1 - k_2$ ($l_1 - l_2$) plane for 
{\it Category} $A$ ($B$) 
get 
slightly extended. Moreover, one finds that $\theta_{23}^\lambda \leq 45^o$ 
as well as  $0\leq \theta_{13}^\lambda \leq 2.7^o$ for {\it Category} $A$ and 
$45^o \leq \theta_{23}^\lambda$
as well as  $0\leq \theta_{13}^\lambda \leq 0.85^o$ for {\it Category} $B$. The 
upper bounds on $\theta_{13}^\lambda$ in both categories correspond to 
$\tan\beta = 60$.
\vskip 0.1in
\n
RG evolution from $\Lambda$ to $\lambda$ has no direct effect on the baryon 
asymmetry $\eta$. The lepton asymmetry $Y_l$, produced at the heavy Majorana 
neutrino mass scale, remains frozen till the temperature comes down to the 
weak scale where it is converted to $\eta$. The requirement of the latter being in the observed range leads to correlated constraints on $x$, $M_{lowest}$ and 
$\tan\beta$, vide tables 1 -- 3. While the constraints on $x$ and $M_{lowest}$ 
have some effects on the magnitude of $\Lambda$, they are numerically quite 
weak. Such is, however, not the case with the $\tan\beta$ constraints, owing 
to eq. (6.2). In particular, the bounds on $\theta_{13}^\lambda$ can be 
significantly affected by restrictions on $\tan\beta$.
\vskip 0.1in
\n
Let us discuss the consequent effects on the said bounds in the three regimes.
\vskip 0.1in
\n
(1) Flavor independent leptogenesis. Here $\tan\beta$ can go from 2 to 60, 
as taken in Ref.[15], for the NHN and QDN cases of {\it Category} $A$ 
and the QDN (NON) 
case of {\it Category} $B$, cf. Table 1. Therefore the range of 
$\theta_{13}^\lambda$ 
remains unchanged for those cases. 
But the stronger restrictions on $\tan\beta$ given in 
Table 1 for the IHN case of {\it Category} $A$ and the NHN, IHN and QDN 
(ION) cases of {\it Category} $B$ 
force the corresponding $\theta_{13}^\lambda$ and $\theta_{23}^\lambda$ to be 
practically equal to  $0^o$ and $45^o$ respectively for those two situations. 
\vskip 0.1in
\n
(2) Fully flavored leptogenesis. We can deduce from the information given in 
table 2 that the ranges of $\theta_{13}^\lambda$ are affected here for either 
category in each case. The results are given in table 4.

\begin{table}[htb]
\begin{center}
{\small
\begin{tabular}{|l|l|l|l||l|l|l|}
\hline
\hline
& \multicolumn{3}{|c||}{Category $A$} & \multicolumn{3}{|c|}{Category $B$}\\
\hline
\hline
&NHN&IHN&QDN&NHN&IHN&QDN\\
\hline
&&&&&&\\
$\tan\beta$&25-60&$22-60$&$2-60$&$16-60$&$24-60$&$6-60$\,(NON)\\
&&&&&&$7-60$\,(ION)\\
\hline
$\theta_{13}^\lambda$ &$0.47^\circ-2.7^\circ$&$0.36^\circ-2.7^\circ$
&$0^\circ-2.7^\circ$&$0.06^\circ-0.85^\circ$&
$0.14^\circ-0.85^\circ$&$0^\circ-0.85^\circ$\\
\hline
\end{tabular}
\caption{Effect on $\theta_{13}^\lambda$ of the more restricted
range of $\tan\beta$ in fully
flavored leptogenesis.}
}
\end{center}
\end{table}
\vskip 0.1in
\n
(3) $\tau$-flavored leptogenesis. There is no additional restriction on 
$\tan\beta$ here as compared with unflavored leptogenesis, 
vide table 3. Hence the ranges of 
$\theta_{13}^\lambda$ stand unchanged in either category for the NHN, IHN 
and QDN cases. 

Now that there is a nonzero $\theta_{13}^\lambda$, one has CP violation 
in the neutrino sector which can be measured from the difference
in oscillation probabilities 
$P(\nu_\mu\rightarrow\nu_e)- P({\bar{\nu_\mu}}\rightarrow{\bar{\nu_e}})$ 
\cite{Minakata}. 
For the CKM CP phase $\delta^{\lambda}$, 
we find the $3\sigma$ range of its value to be
$1.0^o\le \delta^{\lambda}\le 70^o$  
({\it Category} $A$) and $1.5^o\le \delta^{\lambda}\le 90^o$ 
({\it Category} $B$) 
for both flavored and unflavored leptogenesis in all regimes.
 The sign of $\delta^{\lambda}$ is 
opposite to the sign of $\bar\alpha$/$\bar\beta$ 
for {\it Category} $A$/$B$ and hence it does change from one regime for 
$M_{lowest}{(1+\tan^2\beta)}^{-1}$ to another for a given mass ordering 
of $N_i$.
\section{Conclusion}
\n
In this paper we have studied the generation of the observed amount of 
baryon asymmetry $\eta$ in our scheme of $\mu\tau$ symmetric four 
zero neutrino Yukawa textures within the type-I seesaw. For each of the two 
categories $A$ and $B$ of our scheme, we have identified three regimes 
depending on the value of $M_{lowest}{(1 + \tan^2\beta)}^{-1}$ and 
have studied the normal-hierarchical (NHN), inverted-hierarchical (IHN) 
and quasidegenerate (QDN) cases for the masses of the heavy Majorana 
neutrinos $N_i$. The requirement of matching the right value of 
$\eta$ forces the phases
$\bar\alpha$ ({\it Category} $A$) and $\bar\beta$ ({\it Category} $B$) 
to be in the 
fourth quadrant for the NHN and QDN cases and in the first quadrant for 
the IHN case in each regime. Restrictions on small but nonzero $\theta_{13}$, 
arising from radiative $\mu\tau$ symmetry breaking, have also been worked out.
\section{Acknowledgments}
{We thank K.~S.~Babu for suggesting this investigation. 
P.~R. has been supported 
by a DAE Raja Ramanna fellowship.}
\vskip 0.1in
\n
{\bf\it Note added} 
\vskip 0.1in
\n
A new paper on supersymmetric leptogenesis appeared 
\cite{note} after this work was completed. The authors of ref.\,\,\cite{note}   
have 
highlighted certain additional contributions to 
$Y_\Delta$. These arise from soft supersymmetry breaking effects involving 
gauginos and higgsinos as well as anomalous 
global symmetries causing a different pattern of sphaleron 
induced lepton flavor mixing. 
While some of the numerical 
coeffcients -- given in the various expressions for $\eta$ in our 
analysis -- are 
likely to change if these effects are included, their overall signs will 
not. Consequently, there will be no alteration in our conclusions on the 
quadrants of the phases 
$\bar\alpha$ and $\bar\beta$ which remain robust.
\newpage
\appendix
\section {Baryon Asymmetry in flavor independent leptogenesis}
\vskip 0.1in
\n
{{\it Category}\,\, $A$} 
\begin{eqnarray}
\eta_A^{NHN}&\simeq &2.47\times {10}^{-10}
\frac{M_1}{10^9 {\rm GeV}}
\frac{k_2^2\sin2\bar\alpha}{X_A^{1/2}\sin^2\beta}
\frac{M_{2=3}}{M_1}f\left(M_{2=3}^2/M_1^2\right)\times
\nonumber\\&&
{\left[
\frac{1.46\sin^2\beta
X_A^{1/2}}{k_1^2}+
{\left(\frac{28.3 k_1^2}{X_A^{1/2}\sin^2\beta}\right)}^{1.16}
\right]}^{-1}.
\end{eqnarray}
\begin{eqnarray}
\eta_A^{IHN}&\simeq &-2.43\times {10}^{-10}
\frac{M_{2=3}}{10^9 {\rm GeV}}
\frac{k_1^2k_2^2\sin2\bar\alpha}{(1+k_2^2)
X_A^{1/2}\sin^2\beta}
\frac{M_{1}}{M_{2=3}}f\left({M_{1}^2}/{M_{2=3}^2}\right)
\times
\nonumber\\&&
{\left[
1.47{(1+k_2^2)}^{-1}
X_A^{1/2}\sin^2\beta+
{\left(\frac{28\,\,(1+k_2^2)}{X_A^{1/2}\sin^2\beta}\right)}
^{1.16}\right]}^{-1}.
\end{eqnarray}
$$
\eta_A^{QDN}\simeq  2.40\times {10}^{-10}
\frac{k_2^2\sin2\bar\alpha}{X_A^{1/2}\sin^2\beta}
\Bigg\{\frac{M_1}{10^9 {\rm GeV}}
\frac{M_{2=3}}{M_1}f\left(M_{2=3}^2/M_1^2\right)\times
$$
$$
{\left[
1.48{(k_1^2)}^{-1}
X_A^{1/2}\sin^2\beta+
{\left(\frac{27.9\,\,k_1^2)}{X_A^{1/2}\sin^2\beta}\right)}^{1.16}\right]}^{-1}
$$
$$
- \frac{M_{2=3}}{10^9 {\rm GeV}}
\frac{k_1^2}{(1+k_2^2)}
\frac{M_{1}}{M_{2=3}}f\left({M_{1}^2}/{M_{2=3}^2}\right)
\times
{\left[
1.48{(1+k_1^2)}^{-1}
X_A^{1/2}\sin^2\beta +
{\left(\frac{27.9\,\,(1+k_2^2)}{
X_A^{1/2}\sin^2\beta}\right)}^{1.16}\right]}^{-1}\Bigg\}.
\eqno({\rm A}.3)
$$
\ff
{{\it Category}\,\, $B$}
\ef
\vskip 0.1in
$$
\eta_B^{NHN}\simeq 
2.47\times {10}^{-10}
\frac{M_1}{{10}^9{\rm{GeV}}}
\frac{l_2^2\sin2\bar\beta}{(l_1^2+2l_2^2)X_B^{1/2}\sin^2\beta}
\frac{M_{2=3}}{M_1}
f\left(
{M_{2=3}^2/M_1^2}
\right)\times
$$$$
{\left[
\frac{1.46 X_B^{1/2}\sin^2\beta}{(l_1^2+2l_2^2)} + 
{\left(
\frac{28.3(l_1^2+2l_2^2)}{X_B^{1/2}\sin^2\beta}
\right)}^{1.16}
\right]}^{-1}.
\eqno({\rm A}.4)
$$
\newpage
$$
\eta_B^{IHN}\simeq 
-2.43\times {10}^{-10}
\frac{M_{2=3}}{{10}^9{\rm{GeV}}}
\frac{l_2^2\sin2\bar\beta}{X_B^{1/2}\sin^2\beta}
f\left(
{M_1^2/M_{2=3}^2}
\right)\times
$$$$
{\left[
1.47 X_B^{1/2}\sin^2\beta + 
{\left(
\frac{28.0}{X_B^{1/2}\sin^2\beta}
\right)}^{1.16}
\right]}^{-1}.
\eqno({\rm A}.5)
$$
$$
\eta_B^{QDN}\simeq 
2.40\times {10}^{-10}
\Bigg\{\frac{M_1}{{10}^9{\rm{GeV}}}
\frac{l_2^2\sin2\bar\beta}{(l_1^2+2l_2^2)X_B^{1/2}\sin^2\beta}
\frac{M_{2=3}}{M_1}
f\left(
{M_{2=3}^2/M_1^2}
\right)\times
$$$$
{\left[
\frac{1.48 X_B^{1/2}\sin^2\beta}{(l_1^2+2l_2^2)} + 
{\left(
\frac{27.9(l_1^2+2l_2^2)}{X_B^{1/2}\sin^2\beta}
\right)}^{1.16}
\right]}^{-1}
$$
$$
-\frac{M_{2=3}}{{10}^9{\rm{GeV}}}
\frac{l_2^2\sin2\bar\beta}{X_B^{1/2}\sin^2\beta}
f\left(
{M_1^2/M_{2=3}^2}
\right)\times$$$$
{\left[
1.48 X_B^{1/2}\sin^2\beta + 
{\left(
\frac{27.9}{X_B^{1/2}\sin^2\beta}
\right)}^{1.16}
\right]}^{-1}\Bigg\}.
\eqno({\rm A}.6)
$$
\section{Baryon Asymmetry in fully flavored leptogenesis}
\vskip 0.1in
\n
{{\it Category}\,\, $A$}
$$
\eta_A^{NHN}\simeq 
2.47\times {10}^{-10}\frac{M_1}{10^9 {\rm GeV}}
\frac{k_2^2\sin2\bar\alpha}{X_A^{1/2}\sin^2\beta}
\frac{M_{2=3}}{M_1}f\left(M_{2=3}^2/M_1^2\right)
\times$$$$
{\left[
\frac{1.72 X_A^{1/2}\sin^2\beta}{k_1^2}
+{\left(
\frac{23.9k_1^2}{X_A^{1/2}\sin^2\beta}
\right)}^{1.16}
\right]}^{-1}.
\eqno({\rm B}.1)
$$
$$
\eta_A^{IHN}\simeq 
-2.43\times {10}^{-10}
\frac{M_{2=3}}{10^9 {\rm GeV}}
\frac{k_1^2k_2^2\sin2\bar\alpha}{(1+k_2^2)
X_A^{1/2}\sin^2\beta}
f\left({M_{1}^2}/{M_{2=3}^2}\right)
\times $$$$
{\left[
\frac{1.74 X_A^{1/2}\sin^2\beta}{k_2^2}
+{\left(
\frac{23.8k_2^2}{X_A^{1/2}\sin^2\beta}
\right)}^{1.16}
\right]}^{-1}.
\eqno({\rm B}.2)
$$
\newpage
$$
\eta_A^{QDN} \simeq 
2.40\times {10}^{-10}
\Bigg\{\frac{M_1}{10^9 {\rm GeV}}
\frac{k_2^2\sin2\bar\alpha}{X_A^{1/2}\sin^2\beta}
\frac{M_{2=3}}{M_1}f\left(M_{2=3}^2/M_1^2\right)
\times $$$$
{\left[
\frac{1.75 X_A^{1/2}\sin^2\beta}{k_1^2}
+{\left(
\frac{23.6k_1^2}{X_A^{1/2}\sin^2\beta}
\right)}^{1.16}
\right]}^{-1}
$$
$$
-\frac{M_{2=3}}{10^9 {\rm GeV}}
\frac{k_1^2k_2^2\sin2\bar\alpha}{(1+k_2^2)
X_A^{1/2}\sin^2\beta}
f\left({M_{1}^2}/{M_{2=3}^2}\right)
\times 
$$$$
{\left[
\frac{1.75 X_A^{1/2}\sin^2\beta}{k_2^2}
+{\left(
\frac{23.6k_2^2}{X_A^{1/2}\sin^2\beta}
\right)}^{1.16}
\right]}^{-1}\Bigg\}.
\eqno({\rm B}.3)
$$
\n
{{\it Category}\,\, $B$}
$$
\eta_B^{NHN}\simeq 
2.47\times {10}^{-10}
\frac{M_1}{{10}^9{\rm{GeV}}}
\frac{l_2^2\sin2\bar\beta}{(l_1^2+2l_2^2)X_B^{1/2}\sin^2\beta}
\frac{M_{2=3}}{M_1}
f\left(
{M_{2=3}^2/M_1^2}
\right)\times $$$$
{\left[
\frac{2.30 X_B^{1/2}\sin^2\beta}{l_2^2}
+{\left(
\frac{17.9l_2^2}{X_B^{1/2}\sin^2\beta}
\right)}^{1.16}
\right]}^{-1}.
\eqno({\rm B}.4)
$$
$$
\eta_B^{IHN}\simeq 
-2.43\times {10}^{-10}
\frac{M_{2=3}}{{10}^9{\rm{GeV}}}
\frac{l_2^2\sin2\bar\beta}{X_B^{1/2}\sin^2\beta}
\frac{M_1}{M_{2=3}}
f\left(
{M_1^2/M_{2=3}^2}
\right)\times
$$$$
{\left[
{2.32 X_B^{1/2}\sin^2\beta}
+{\left(
\frac{17.8}{X_B^{1/2}\sin^2\beta}
\right)}^{1.16}
\right]}^{-1}.
\eqno({\rm B}.5)
$$
$$
\eta_B^{QDN} \simeq 
2.40\times {10}^{-10}
\Bigg\{\frac{M_1}{{10}^9{\rm{GeV}}}
\frac{l_2^2\sin2\bar\beta}{(l_1^2+2l_2^2)X_B^{1/2}\sin^2\beta}
\frac{M_{2=3}}{M_1}
f\left(
{M_{2=3}^2/M_1^2}
\right)\times
$$$$
{\left[
\frac{2.34 X_B^{1/2}\sin^2\beta}{l_2^2}
+{\left(
\frac{17.7l_2^2}{X_B^{1/2}\sin^2\beta}
\right)}^{1.16}
\right]}^{-1}
-\frac{M_{2=3}}{{10}^9{\rm{GeV}}}
\frac{l_2^2\sin2\bar\beta}{X_B^{1/2}\sin^2\beta}
\frac{M_1}{M_{2=3}}
f\left(
{M_1^2/M_{2=3}^2}
\right)\times
$$$$
{\left[
{2.34 X_B^{1/2}\sin^2\beta}
+{\left(
\frac{17.7}{X_B^{1/2}\sin^2\beta}
\right)}^{1.16}
\right]}^{-1}\Bigg\}.
\eqno({\rm B}.6)
$$
\newpage
\section{Baryon asymmetry in  $\tau$-flavored leptogenesis }
\n
{{\it Category}\,\, $A$}
$$
\eta_A^{NHN}\simeq 
2.47\times {10}^{-10}
\frac{M_1}{10^9 {\rm GeV}}
\frac{k_2^2\sin2\bar\alpha}{X_A^{1/2}\sin^2\beta}
\frac{M_{2=3}}{M_1}f\left(M_{2=3}^2/M_1^2\right)
\times $$$$
{\left[
\frac{2.05 X_A^{1/2}\sin^2\beta}{k_1^2}
+{\left(
\frac{20.1k_1^2}{X_A^{1/2}\sin^2\beta}
\right)}^{1.16}
\right]}^{-1}.
\eqno({\rm C}.1)
$$
$$
\eta_A^{IHN}\simeq 
-1.22\times {10}^{-10}\frac{M_{2=3}}{10^9 {\rm GeV}}
\frac{k_1^2k_2^2\sin2\bar\alpha}{(1+k_2^2)
X_A^{1/2}\sin^2\beta}
\frac{M_{1}}{M_{2=3}}f\left({M_{1}^2}/{M_{2=3}^2}\right)
\times $$$$\left\{
{\left[
\frac{2.07 X_A^{1/2}\sin^2\beta}{k_2^2}
+{\left(
\frac{20.0k_2^2}{X_A^{1/2}\sin^2\beta}
\right)}^{1.16}
\right]}^{-1}
+
{\left[
\frac{2.07 X_A^{1/2}\sin^2\beta}{(1+k_2^2)}
+{\left(
\frac{20.0(1+k_2^2)}{X_A^{1/2}\sin^2\beta}
\right)}^{1.16}
\right]}^{-1}\right\}.
\eqno({\rm C}.2)
$$
$$
\eta_A^{QDN}\simeq 
1.20\times {10}^{-10}\frac{k_2^2\sin2\bar\alpha}
{X_A^{1/2}\sin^2\beta}\Bigg\{
     \frac{2M_1}{10^9 {\rm GeV}}
\frac{M_{2=3}}{M_1}f\left(M_{2=3}^2/M_1^2\right)
\times $$$$
{\left[
\frac{2.08 X_A^{1/2}\sin^2\beta}{k_1^2}
+{\left(
\frac{19.8k_1^2}{X_A^{1/2}\sin^2\beta}
\right)}^{1.16}
\right]}^{-1}
-\frac{M_{2=3}}{10^9 {\rm GeV}}
\frac{k_1^2}{(1+k_2^2)}
\frac{M_{1}}{M_{2=3}}f\left({M_{1}^2}/{M_{2=3}^2}\right)
\times $$$$\left\{
{\left[
\frac{2.08 X_A^{1/2}\sin^2\beta}{k_2^2}
+{\left(
\frac{19.8k_2^2}{X_A^{1/2}\sin^2\beta}
\right)}^{1.16}
\right]}^{-1}
 + 
{\left[
\frac{2.08\sin^2\beta X_A^{1/2}}{(1+k_2^2)}
+{\left(
\frac{19.8{(1+k_2^2)}}{X_A^{1/2}\sin^2\beta}
\right)}^{1.16}
\right]}^{-1}\right\}\Bigg\}.
\eqno({\rm C}.3)
$$
\n
{{\it Category}\,\, $B$}
$$
\eta_B^{NHN}\simeq 
1.23\times {10}^{-10}\times
\frac{M_1}{{10}^9{\rm{GeV}}}
\frac{l_2^2\sin2\bar\beta}{(l_1^2+2l_2^2)X_B^{1/2}\sin^2\beta}
\frac{M_{2=3}}{M_1}
f\left(
{M_{2=3}^2/M_1^2}
\right)\times
$$
$$\left\{
{\left[
\frac{2.05 X_B^{1/2}\sin^2\beta}{(l_1^2+l_2^2)}
+{\left(
\frac{20.1(l_1^2+l_2^2)}{X_B^{1/2}\sin^2\beta}
\right)}^{1.16}
\right]}^{-1}
+
{\left[
\frac{2.24 X_B^{1/2}\sin^2\beta}{l_2^2}
+{\left(
\frac{18.4l_2^2}{X_B^{1/2}\sin^2\beta}
\right)}^{1.16}
\right]}^{-1}\right\}.
\eqno({\rm C}.4)
$$
$$
\eta_B^{IHN}\simeq 
-1.22\times {10}^{-10}
\frac{M_{2=3}}{{10}^9{\rm{GeV}}}
\frac{l_2^2\sin2\bar\beta}{X_B^{1/2}\sin^2\beta}
\frac{M_1}{M_{2=3}}
f\left(
{M_1^2/M_{2=3}^2}
\right)\times
$$$$
\left\{
{\left[
\frac{2.07 X_B^{1/2}\sin^2\beta}{l_2^2}
+{\left(
\frac{20.0l_2^2}{X_B^{1/2}\sin^2\beta}
\right)}^{1.16}
\right]}^{-1}
 + 
{\left[
2.26 X_B^{1/2}\sin^2\beta
+{\left(
\frac{18.2}{X_B^{1/2}\sin^2\beta}
\right)}^{1.16}
\right]}^{-1}\right\}.
\eqno({\rm C}.5)
$$
\newpage
$$
\eta_B^{QDN}\simeq 
1.20\times {10}^{-10}\frac{l_2^2\sin2\bar\beta}{X_B^{1/2}\sin^2\beta}
\Bigg(
\frac{M_1}{{10}^9{\rm{GeV}}}
\frac{1}{(l_1^2+2l_2^2)}
\frac{M_{2=3}}{M_1}
f\left(
{M_{2=3}^2/M_1^2}
\right)\times $$
$$
\left\{{\left[
\frac{2.16 X_B^{1/2}\sin^2\beta}{(l_1^2+l_2^2)}
+{\left(
\frac{19.1(l_1^2+l_2^2)}{X_B^{1/2}\sin^2\beta}
\right)}^{1.16}
\right]}^{-1} +
{\left[
\frac{2.28 X_B^{1/2}\sin^2\beta}{l_2^2}
+{\left(
\frac{18.1l_2^2}{X_B^{1/2}\sin^2\beta}
\right)}^{1.16}
\right]}^{-1}
\right\}
$$
$$
-\frac{M_{2=3}}{{10}^9{\rm{GeV}}}
\frac{M_1}{M_{2=3}}
f\left(
{M_1^2/M_{2=3}^2}
\right)\times
$$$$
\left\{
{\left[
\frac{2.16 X_B^{1/2}\sin^2\beta}{l_2^2}
+{\left(
\frac{19.1l_2^2}{ X_B^{1/2}\sin^2\beta}
\right)}^{1.16}
\right]}^{-1}+
{\left[
2.28 X_B^{1/2}\sin^2\beta
+{\left(
\frac{18.1}{X_B^{1/2}\sin^2\beta}
\right)}^{1.16}
\right]}^{-1}\right\} \Bigg ).
\eqno({\rm C}.6)
$$

\end{document}